\documentclass[twocolumn,twocolappendix]{aastex63}

\usepackage{xspace}

\newcommand{\kms}{km\,s$^{-1}$\xspace}
\newcommand{\msun}{\ensuremath{M_\odot}\xspace}
\newcommand{\jybm}{Jy\,beam$^{-1}$\xspace}

\accepted{May 22, 2020}
\shorttitle{IRAS 16293 A: orbital and mass constraints }
\shortauthors{Maureira et al.}
\graphicspath{{./}{figures/}}

\usepackage{hyperref}
\hypersetup{linkcolor=blue,citecolor=blue,filecolor=cyan,urlcolor=blue}

\begin{document}

\title{Orbital and mass constraints of the young binary system IRAS 16293-2422 A}

\correspondingauthor{Mar\'ia Jos\'e Maureira}
\email{maureira@mpe.mpg.de}


\author[0000-0002-7026-8163]{Mar\'ia Jos\'e Maureira}
\affiliation{Max-Planck-Institut für extraterrestrische Physik (MPE), Gießenbachstr. 1, D-85741 Garching, Germany}

\author[0000-0002-3972-1978]{Jaime E. Pineda}
\affiliation{Max-Planck-Institut für extraterrestrische Physik (MPE), Gießenbachstr. 1, D-85741 Garching, Germany}

\author[0000-0003-3172-6763]{Dominique M. Segura-Cox}
\affiliation{Max-Planck-Institut für extraterrestrische Physik (MPE), Gießenbachstr. 1, D-85741 Garching, Germany}

\author[0000-0003-1481-7911]{Paola Caselli}
\affiliation{Max-Planck-Institut für extraterrestrische Physik (MPE), Gießenbachstr. 1, D-85741 Garching, Germany}

\author[0000-0003-1859-3070]{Leonardo Testi}
\affiliation{European Southern Observatory, Karl-Schwarzschild-Strasse 2, 85748 Garching, Germany}

\author[0000-0002-2357-7692]{Giuseppe Lodato}
\affiliation{Dipartimento di Fisica, Università degli Studi di Milano, Via Celoria 16, I-20133 Milano, Italy}

\author[0000-0002-5635-3345]{Laurent Loinard}
\affiliation{Instituto de Radioastronomía y Astrofísica, Universidad Nacional Aut\'onoma de M\'exico Morelia, 58089, México}
\affiliation{Instituto de Astronomía, Universidad Nacional Autónoma de México,\\ Apartado Postal 70-264, Ciudad de México 04510, México}

\author[0000-0001-7520-4305]{Antonio Hern\'andez-G\'omez}
\affiliation{Instituto de Radioastronomía y Astrofísica, Universidad Nacional Aut\'onoma de M\'exico Morelia, 58089, México}
\affiliation{Max-Planck-Institute für Radioastronomie Auf dem Hügel 69, 53121 Bonn, Germany}
\affiliation{IRAP, Université de Toulouse CNRS, UPS, CNES, Toulouse, France}



\begin{abstract}

We present 3 mm ALMA continuum and line observations at resolutions of 6.5 au and 13 au respectively, toward the Class 0 system IRAS 16293-2422 A. The continuum observations reveal two compact sources towards IRAS 16293-2422 A, coinciding with compact ionized gas emission previously observed at radio wavelengths (A1 and A2), confirming the long-known radio sources as protostellar. The emission towards A2 is resolved and traces a dust disk with a FWHM size of $\sim$12 au, while the emission towards A1 sets a limit to the FWHM size of the dust disk of $\sim$4 au. We also detect spatially resolved molecular kinematic tracers near the protostellar disks. Several lines of the $J=5-4$ rotational transition of HNCO, NH$_2$CHO and t-HCOOH are detected, with which we derived individual line-of-sight velocities. Using these together with the CS ($J=2-1$), we fit Keplerian profiles towards the individual compact sources and derive masses of the central protostars. The kinematic analysis indicates that A1 and A2 are a bound binary system. Using this new context for the previous 30 years of VLA observations, we fit orbital parameters to the relative motion between A1 and A2 and find the combined protostellar mass derived from the orbit is consistent with the masses derived from the gas kinematics. Both estimations indicate masses consistently higher ($0.5\lesssim M_1\lesssim M_2 \lesssim2$ \msun) than previous estimations using lower resolution observations of the gas kinematics. The ALMA high-resolution data provides a unique insight into the gas kinematics and masses of a young deeply embedded bound binary system.

\end{abstract}

\keywords{Protostars--- 
Close binary stars --- circumstellar dust --- circumstellar gas}


\section{Introduction} \label{sec:intro}

IRAS 16293-2422 (hereafter IRAS 16293) is a well-known bright ($L_{bol}\approx21$ L$_{\odot}$, \citealt{2018JacobsenALMAPILS}) low-mass protostellar system located in the Ophiuchus molecular cloud at a distance of 141 pc \citep{2018DzibRevised}. It has been widely studied because it was one of the first protostellar systems consistent with being in the Class 0 stage \citep{1993AndreSub} and also because of its active 'hot corino' chemistry with numerous complex molecules tracing compact regions with a high excitation ($\gtrsim$ 100 K) temperature \citep{2000ceccarelliStructure,2011CauxTIMASSS,2016JorgensenPils}. Further, IRAS 16293 was one of the first Class 0 sources recognized as a multiple system. It was first resolved into two objects in cm observations showing source A to the south and source B to the north, separated by $\sim$700 au or 5" \citep{1989WoottenDuplicity}. Later, \cite{1992MundyIRAS16293} confirmed the protostellar nature of A and B by detecting compact dust thermal emission at 3 mm, coincident with the location of the cm sources. Source B shows a single peak from cm to submm wavelengths, and it is believed to be at a very early stage of evolution \citep{2012PinedaFirst,2016JorgensenPils,2019HernandezGomezNature}. On the other hand, \cite{1989WoottenDuplicity} 2 cm VLA observations revealed early on two compact sources, A1 and A2 within A, separated by $\sim$ 50 au and aligned approximately perpendicular to the line connecting A and B. Later, \cite{2005ChandlerIRAS16293} also resolved source A into two sub-mm peaks named Ab and Aa separated by $\sim$90 AU and aligned similar to A1-A2. The peak of Aa was located in between A1 and A2. Similar results were obtained with ALMA 1.3 mm observations with a resolution of 0.25" or 35 au \citep{2018SadavoyDustIRAS16293}. By imaging the longest baselines in their observations, they recovered peaks consistent with the sub-mm peaks Ab and Aa identified in \cite{2005ChandlerIRAS16293}, with an additional weaker peak next to Aa, named Aa*. The lack of clear correspondence between the sub-mm and cm sources within source A prevented the confirmation of the nature of A1 and A2 as protostellar. For instance, \cite{2005ChandlerIRAS16293} argued that A1 was a shock feature due to a precessing jet. This claim was based on the large proper motions of A1 with respect to A2 \citep{2002LoinardLarge}, and the shift in P.A. of A1 with respect to A2. On the other hand, \cite{2019HernandezGomezNature} recently argued that A1 is the location of a protostar, due to the nearly constant flux of A1 over time, inconsistent with the expectation of a shock feature. Despite this debate, the observation of possibly three outflows powered within A \citep{1990MizunoRemarkable,2004StarkProbing,2014GirartOrigin,2019VanderWielALMAPILS}, further supported the multiple protostellar nature of source A. \\

Sources A and B are embedded within a dense core with a mass of 4-6 \msun enclosed within several 1,000 au \citep{2018JacobsenALMAPILS,2020LadjelateHerschel}. Source B has a face-on configuration and its mass has been constrained to few 0.1 \msun up to about 1 \msun\ using interferometric observations at a resolution of 70-140 au \citep{2012PinedaFirst,2018OyaSourceB,2018JacobsenALMAPILS}. Source A shows a fattened disk-like structure with a radius of about 100 au in observations at a resolution of 30-140 au \citep{2016JorgensenPils,2018SadavoyDustIRAS16293}. The same observations reveal a velocity gradient along the major axis of the  disk-like structure and the velocity profile was used to constrain the mass for source A, resulting in values around 1 \msun \citep{2016OyaInfalling,2018JacobsenALMAPILS}. On the other hand \cite{2007LoinardNew} estimated a mass of 2-3 \msun\ from assuming that A1 and A2 were two protostars whose relative motion was a circular orbit in the plane of the sky. \\

Here, we present ALMA Band 3 continuum observations with a resolution of 0.046" (6.5 au) that reveal for the first time two compact sources at wavelengths tracing dust thermal emission, coincident with the location of the cm compact sources A1 and A2, thus confirming IRAS 16293 A1-A2 as a binary. Further, we present 3 mm line emission at a resolution of 13 au used to study the gas kinematics. This paper is organized as follows: In Section~\ref{sec:obs} we describe the observations and data reduction. In Section~\ref{sec:results} we present the results of the continuum and line analysis. In Section~\ref{sec:discussion} we discuss the implications of our results and analyze the positions of the new ALMA epoch and previous VLA observations and derived orbital parameters. Section~\ref{sec:sum_conclusions} corresponds to the summary and conclusions.


\section{Observations} \label{sec:obs}

IRAS 16293 (A and B) was observed on October 8 and 12, 2017 using the most extended Cycle 5 configuration of ALMA (41.4 m - 16.2 km baseline range) in Band 3 (single pointing) with a total time on-source of 1.25 hours. The maximum recoverable scale is $\sim$0.5" (or 70 au)\footnote{From the ALMA Cycle 5 proposer's guide. We note that this value is close to the one obtained using 0.983$\times$ (wavelength/$L_5$), with $L_5$ the 5th percentile of uv distance in the observations. This is an empirically determined relation published in the Cycle 8 proposer's guide.} and the pointing center of the observations was ICRS 16:32:22.63 -24:28:31.8. The bandpass/flux calibrator and phase calibrator were J1517-2422 and J1625-2527, respectively. The observations were part of the cycle 5 project ID:2017.1.01247.S. (PI: G. Dipierro). 

The spectral setup consisted of one continuum window centered at 99.988 GHz with 128 channels and a total bandwidth of 2 GHz and four windows of 960, 1920, 960 and 1920 channels with widths of 22.070 kHz centered at the frequencies of $^{13}$CO (1-0), C$^{17}$O (1-0),  C$^{18}$O (1-0), and CS (2-1), respectively. CASA 5.4.1 \citep{2007McMullinCASA} was used for both calibration and imaging. Calibration of the raw visibility data was done using the standard pipeline. When imaging the continuum we iteratively performed phase-only self-calibration with a minimum solution interval of 9 seconds. Afterwards we performed two amplitude self-calibration iterations, with a minimum solution interval of 60 seconds. The final continuum dataset after phase+amplitude self-calibration was imaged using the tclean task with the multiscale deconvolver and a robust parameter of 0.5. We tried different scales for the multiscale imaging and for the final image we adopted those that resulted in the minimal residuals and no significant artifacts. The adopted values are four scales of 0, 8, 24 and 72 pixels (pixel size of 6 mas). These scales correspond to approximately 0 (point source), 1, 3 and 9 times the beam size. The beam size, beam P.A. and noise of the final continuum image are 0.048"$\times$0.046" (6.5 au), 79.3$^{\circ}$ and 15 $\mu$\jybm, respectively. 

The continuum self-calibration solutions were applied to all four line windows, 
to improve the signal-to-noise ratio \citep{2018BroganAdvanced}. Continuum subtraction was performed with CASA task uvcontsub by selecting line-free channels. We identified CS ($J=2-1$) and several lines of the $J=5-4$ rotational transition of HNCO, NH$_2$CHO and t-HCOOH. These, along with $^{13}$CO (1-0) were the brightest emission lines in the data. One extra bright line was also observed but due to uncertainties with their identification, we do not use it for our analysis and is not shown here. C$^{17}$O (1-0) was undetected and C$^{18}$O (1-0) showed only extended, very weak emission and thus was not used for the analysis. We cleaned CS (2-1), the $J=5-4$ rotational transition of HNCO, NH$_2$CHO and t-HCOOH, and $^{13}$CO (1-0) cubes using natural weighting, the multiscale deconvolver with scales of 0, 5 and 15 pixels, a pixel size of 0.018”, a channel width of 0.38 \kms and a uvtaper parameter of 0.06" to better recover moderately extended features. Those scales correspond to 0 (point source), 1 and 3 times the beam. As with the continuum image, we picked those scales based on examination of the  residuals. The cube of CS (2-1) and $^{13}$CO (1-0) show strong imaging artifacts (e.g. stripes) due to missing flux in the velocity range 3.24-5.9 \kms (VLSR of the large-scale cloud). For t-HCOOH we produced an average cube using four sub-levels of the $J=5-4$ transitions that were isolated. Appendix Table~\ref{tb:lines} summarizes the lines used with their corresponding levels, sub-levels, frequencies and upper level energies. The average final beam size, beam P.A., and noise for the lines are 0.106"$\times$0.084", 55.7$^{\circ}$ and 1.1 mJy/beam per channel. 

\section{Results}
\label{sec:results}
\subsection{ALMA 3 mm continuum emission} \label{sec:cont}

\begin{figure*}[ht!]
\includegraphics[scale=1.1]{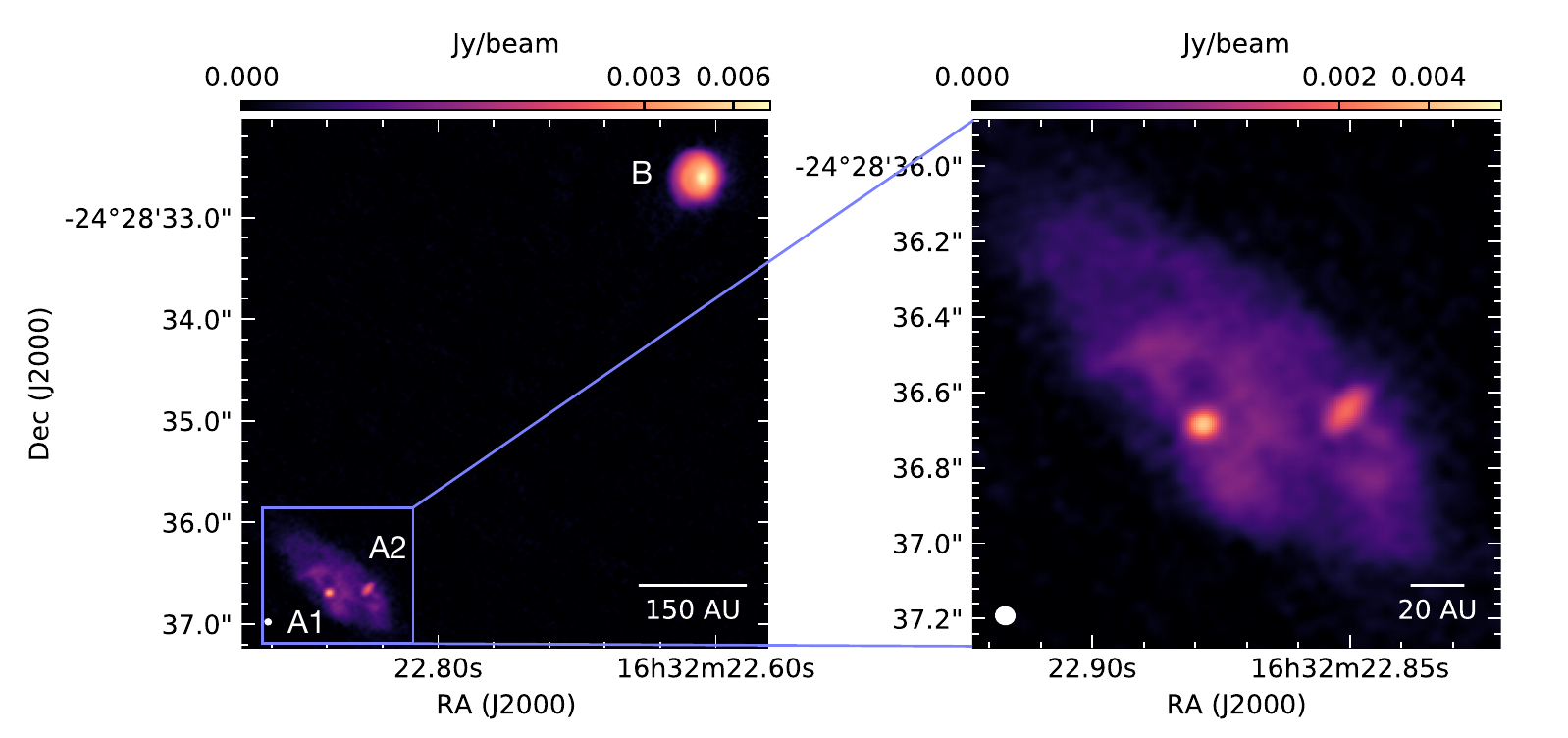}
\caption{ALMA High angular resolution images of the multiple protostellar system IRAS 16293. Left: 3 mm continuum observations showing the triple nature of IRAS16293. Source B to the north, hosts a single protostar, embedded within a 40 au across close to face-on disk-like structure.
The flux from the two compact 3 mm sources to the south remain significant after removing contamination from free-free, unambiguously confirming the radio sources A1 and A2 as a binary protostellar system. Right:  Zoom-in view towards source A. The bright and compact (major axis $<$12 au) sources, separated by 54 au, likely correspond to two individual circumstellar disks. Extended dust structures surrounding the circumstellar disks are also revealed. \label{fig:3mm}}
\end{figure*}

Figure~\ref{fig:3mm} shows the high-resolution 3 mm ALMA observations towards IRAS 16293. The 3 mm counterparts of the radio sources A1 and A2 are clearly detected. The compact emission towards A2 is resolved into an elongated disk-like structure while the compact and brighter emission towards A1 appears unresolved. The projected separation of A1 and A2  is 0.38" or 54 au. The further improvement in sensitivity achieved through self-calibration resulted in further detection of weaker extended structures. A circumbinary disk-like structure with a semi-major axis of about $\sim$0.7" (100 au) and P.A.$\sim50^{\circ}$ is observed, in agreement with previous lower resolution observations \citep{2016JorgensenPils,2016OyaInfalling,2018SadavoyDustIRAS16293}. Overlaid on the circumbinary disk-like structure there are also newly seen complex narrower features.\\

We performed 2D Gaussian fits on the image plane of the bright and compact continuum emission towards A1 and A2. We report here the results of fits with background subtraction since the residuals for this fit compared with one without background subtraction are substantially better (Figure~\ref{fig:cont_maps_fit}). See Appendix \ref{ap:contfit} for details of the procedure and comparisons. Table~\ref{tb:3mmfit} lists the results of the fit. Using the sizes in Table~\ref{tb:3mmfit}, the inferred inclinations assuming circular geometry are 59$^{\circ}$ $\pm$ 4 and 74$^{\circ}$ $\pm$ 1 from the plane-of-sky for A1 and A2, respectively. To obtain an estimation of the inclination of the circumbinary disk-like structure we performed another 2D Gaussian fit. For this, we use 3 components, one for each of the compact emission, A1 and A2, and one for the extended structure. Assuming circular geometry we inferred an inclination of 64$^{\circ}$ $\pm$ 1 for the circumbinary disk-like structure with a P.A. of 50$^{\circ} \pm 1$. Although the inclinations inferred for both of the compact sources and the extended material agree within 10$^{\circ}$, their P.A. are misaligned. The P.A. of the compact resolved emission towards A2 is about 138$^{\circ}$, resulting in an almost $\sim90^{\circ}$ misalignment with that of the circumbinary disk-like structure (P.A.$\sim$50$^{\circ}$). Although a similar misalignment is derived for A1 according to the results from the fit, the compact A1 emission is unresolved making this measurement uncertain. \\

\begin{deluxetable*}{lcccccc}[ht]
\tablecaption{Fit to the 3mm compact sources A1 and A2}             
\label{tb:3mmfit}      
\centering                          
\tablecolumns{7}
\tablewidth{0pt}
\tablehead{
\colhead{Source} & \colhead{R.A.} & \colhead{Decl.} & \colhead{Deconvolved Size} & 
\colhead{P.A.} & \colhead{Peak Flux Density} & \colhead{Integrated Flux}\\
\colhead{} & \colhead{(J2000)} & \colhead{(J2000)} & \colhead{(marcsec)} & 
\colhead{($^\circ$)} & \colhead{(mJy beam$^{-1}$)} & \colhead{(mJy)}
}
\startdata
A1&16:32:22.878 & -24:28:36.684&(24.6 $\pm$ 0.8 , 12.8 $\pm$ 1.2)  &  119.8 $\pm$ 3.6&4.18 $\pm$ 0.02&4.91 $\pm$ 0.04 \\
A2&16:32:22.851 & -24:28:36.647&(87.3 $\pm$ 2.5 , 23.7 $\pm$ 2.0)  &137.9 $\pm$ 1.0 & 1.73 $\pm$ 0.03 &4.23 $\pm$ 0.11
\enddata
\end{deluxetable*}

In the context of previous radio and sub-mm observations, it is likely that the two compact emission come from small circumstellar disks. Previous VLA observations provide a spectral index at the low-frequency end, which for source A1 and A2 are consistent with free-free emission from an ionized jet. \cite{2019HernandezGomezNature} reported a spectral index from VLA observations of $0.5\pm0.2$ and $0.7\pm0.2$ for A1 and A2, respectively. From the most recent 7 mm observations with the VLA (epoch 2013) where A1 and A2 are clearly resolved \citep{2019HernandezGomezNature}, we estimated fluxes of 1.83$\pm$0.08 and 2.24$\pm$0.13 mJy for A1 and A2 at 3 mm. This results in free-free contamination of $37\pm2$\% and $53\pm3$\% for A1 and A2, respectively. These are conservative upper limits estimates of the free-free fluxes since attempts to fit the compact sources with the sum of a point source and a Gaussian results in even less flux coming from the point source (taken as the unresolved free-free) than the above extrapolation (Appendix Section~\ref{ap:contfit}). Thus, at least half of the flux in both sources comes from thermal dust emission tracing the location of two protostars. 

The 3 mm compact emission towards A2 is perpendicular to the bipolar ejecta observed at cm wavelengths within 100 au \citep{2007LoinardNew,2010PechConfirmation}. Recent $\sim$100 au resolution water maser observations also show blueshifted emission from A2, moving along the bipolar ejecta direction \citep{2018DzibRevised}. The bipolar ejecta is also aligned with a 0.1 pc scale CO molecular outflow \citep{1990MizunoRemarkable,2004StarkProbing}, but this molecular outflow has no clear counterpart below $\sim$700 au \citep{2008YehCO,2014GirartOrigin,2019VanderWielALMAPILS} and the blueshifted lobe is at the opposite side of the blueshifted water maser emission \citep{2018DzibRevised}. Despite the confusion with the molecular outflow, the evidence from the ejecta and the flux coming from thermal dust emission support that the compact structure revealed in the 3 mm observations towards A2 is tracing a dust circumstellar disk with an approximate size of 12 au (from the FWHM in Table~\ref{tb:3mmfit}). \\

The emission towards A1 is not well resolved. There are no previous observations of ejecta from A1 and this source also has not been unambiguously associated with an individual molecular outflow yet. However, besides the CO outflow mentioned above and possibly related to A2, there are two other CO outflows powered within A \citep{1990MizunoRemarkable,2013KristensenALMACO,2014GirartOrigin,2019VanderWielALMAPILS}; one oriented East-West and one pointing towards B. We note that although it is expected than one of them is being driven by A1, none of these CO outflows is aligned with the inferred (although unresolved) minor axis of the A1 dust disk. Our observations constrain the size of the dust circumstellar disk around A1 to $\lesssim$3.5 au (from the FWHM in Table~\ref{tb:3mmfit}).\\

Finally, we note that the previous super-resolution images at frequencies $>$ 200 GHz  that revealed the peaks Ab and Aa, not matching the location of A1 and A2 (particularly A2, \citealt{2005ChandlerIRAS16293,2013ChenSMA,2018SadavoyDustIRAS16293}) were likely affected by the high optical depth of the surrounding material, which prevented the clear detection of the embedded compact sources. Our observations show that the location of Ab and Ab from \citealt{2018SadavoyDustIRAS16293} are tracing the substructures around A1 while Aa*, the weakest additional peak next to Aa identified in \citep{2018SadavoyDustIRAS16293}, is near A1 (Figure~\ref{fig:chand_peaks}).

\subsubsection{Masses from 3 mm continuum emission}

Gas masses from the 3 mm continuum are commonly estimated using:
\begin{equation}
    M=100\frac{d^2S_{\nu}}{B_{\nu}(T_{d})\kappa_{\nu}}
    \label{eq:massfromdust}
\end{equation}

where $S_{\nu}$ is the integrated flux density, $B_{\nu}$ is the Planck function, $\kappa_{\nu}$ is the dust opacity and $d$ is the distance. Equation~\ref{eq:massfromdust} assumes optically thin emission \citep{1983HIldebrandDetermination}. $S_{\nu}$ is taken from Table~\ref{tb:3mmfit} and corrected by the free-free contamination. We assume a dust temperature of 100 K \citep{2016OyaInfalling,2020VantHoffTemperature} and a gas-to-dust ratio of 100. For the dust opacity we adopt a range of values between 0.23 cm$^2$g$^{-1}$, appropriate for dense material at a very young evolutionary state \citep{1994OssenkopfDust,2016DunhamALMA}, and 0.81 cm$^2$g$^{-1}$, corresponding to dense material at a more evolved state \citep{2019AgurtoGangasRevealing}. For obtaining an estimate of the mass of the extended emission towards Source A, we use the integrated flux of all emission above 3$\sigma$ after subtracting the contribution from the compact sources A1 and A2. We obtained a total integrated flux from the extended emission of $\sim$73 mJy. The range of masses calculated with the two opacities for A1, A2  and the extended emission correspond to $1-3\times10^{-3}$ \msun, $1-3\times10^{-3}$ \msun, and $0.03-0.1$ \msun, respectively. There are important additional uncertainties on these values arising from uncertainties in the dust temperature, optical depth and as previously mentioned, the true contribution from free-free emission. All of these factors contribute independently and can decrease or increase the reported values within a factor of a few \citep{2019BalleringProtoplanetary}. Further, given that at the scales of the observed circumstellar disks dust scattering can decrease the intensity at millimeter wavelengths, which are likely also optically thick, our estimates for the mass of the compact sources should be taken as conservative lower limits \citep{2019LiuAnomalously,2020UedaScattering}. The order of magnitude of the lower limits reported here are comparable to other compact ($\lesssim$40 au) circumstellar disks in Class I multiple systems observed at high ($\lesssim$25 au) resolution, derived in a similar fashion \citep{2017TakakuwaSpiral,2019AlvesGas,2019CruzSaenzResolved}.



\subsection{Molecular lines}
\label{sec:moments}
\begin{figure*}[ht!]
\includegraphics[scale=0.5]{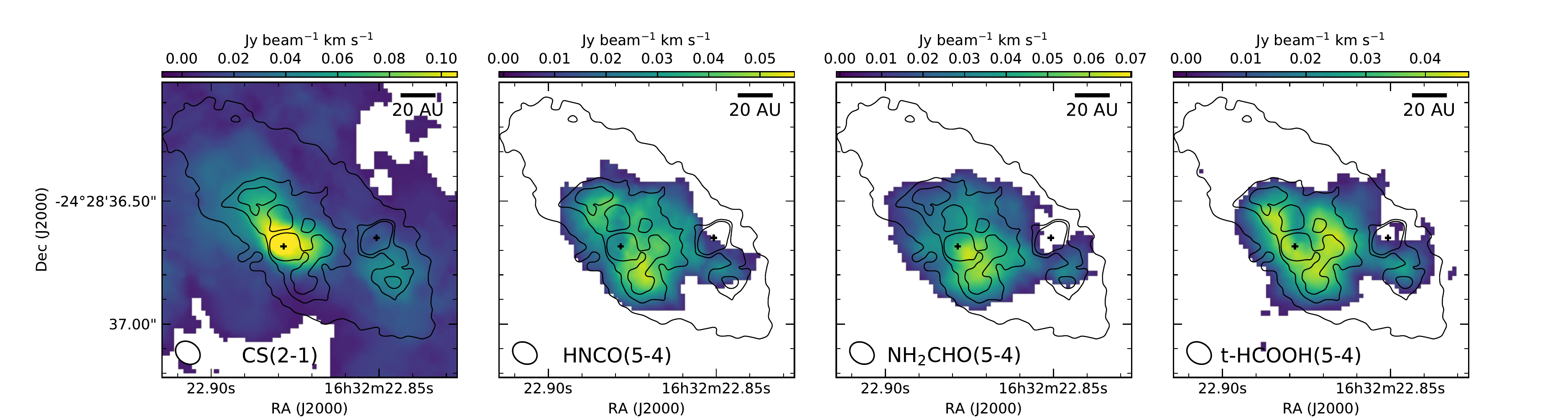}
\caption{Moment 0 maps towards IRAS 16293 A (color).  The velocity range for CS (2-1) was split into two to avoid channels with artifacts due to missing flux. The two ranges were $[-7.02,2.86]$ \kms\ and $[6.28,12.74]$ \kms. For HNCO and t-HCOOH the velocity range was $[-7.02,12.72]$ \kms, while for NH$_2$CHO we restricted the range to $[-0.18,12.74]$ \kms\ since at blueshifted velocities the channels also showed emission from a different line. Black contours show the 3 mm continuum emission at levels 124 $\mu$Jy, 320 $\mu$Jy and 448 $\mu$Jy to identify the circumbinary disk-like structure, the compact sources A1 and A2 and the smaller scales substructures around them. Crosses mark the peak location for A1 and A2 (Table~\ref{tb:3mmfit}).  The beam is shown in the bottom left corner of each panel. 
\label{fig:mom0}}
\end{figure*}

\begin{figure*}[ht!]
\includegraphics[scale=0.49]{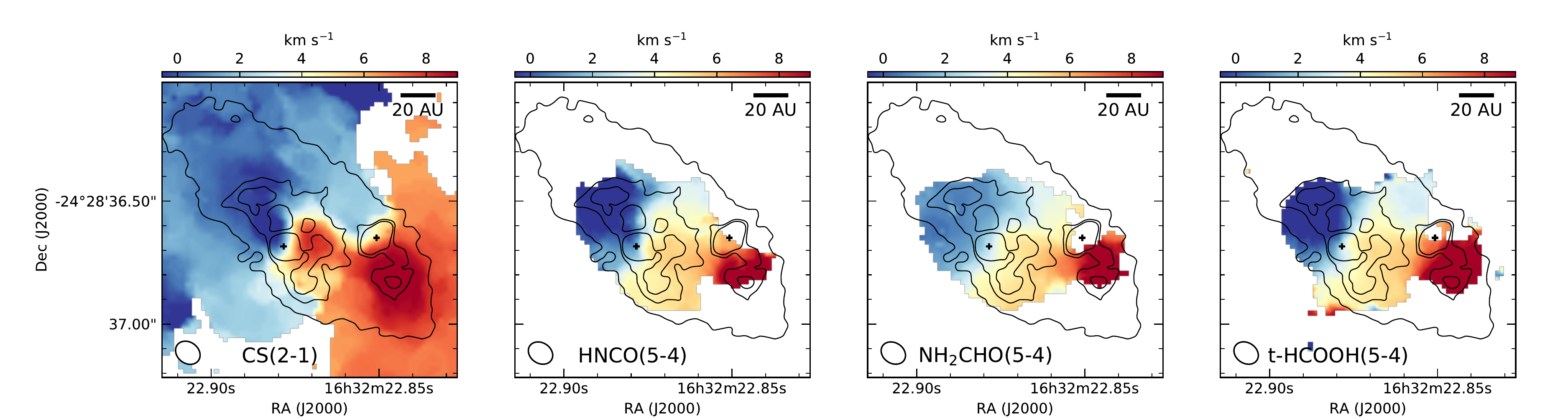}
\caption{Moment 1 maps towards IRAS 16293 A (color). The velocity range for CS (2-1) was split into two to avoid channels with artifacts due to missing flux. The two ranges were $[-7.02,2.86]$ \kms\ and $[6.28,12.74]$ \kms. For HNCO and t-HCOOH the velocity range was $[-7.02,12.72]$ \kms, while for NH$_2$CHO we restricted the range to $[-0.18,12.74]$ since at blueshifted velocities the channels also showed emission from a different line. Black contours show the 3 mm continuum emission at levels 124 $\mu$Jy, 320 $\mu$Jy and 448 $\mu$Jy to identify the circumbinary disk-like structure, the compact sources A1 and A2 and the smaller scales substructures around them. Crosses mark the peak location for A1 and A2 (Table~\ref{tb:3mmfit}). The beam is shown in the bottom left corner of each panel.  \label{fig:mom1}}
\end{figure*}

Figures~\ref{fig:mom0} and~\ref{fig:mom1} show the moment 0 and moment 1 maps for the CS (2-1), HNCO (5-4), NH$_2$CHO (5-4) and t-HCOOH (5-4) molecular lines. The moment 0 and 1 maps are integrated over the same velocity ranges. In addition, only pixels with emission $>$3$\sigma$ per channel were considered. The velocity range for CS was split into two to avoid channels with artifacts (due to missing flux). The two ranges correspond to $[-7.02,2.86]$ \kms\ and $[6.28,12.74]$ \kms. For HNCO and t-HCOOH the velocity range is $[-7.02,12.72]$ \kms, while for NH$_2$CHO we restricted the range to $[-0.18,12.74]$ \kms\ since at blueshifted velocities the channels also showed emission from an unrelated line. \\

The moment 0 maps for HNCO (5-4), NH$_2$CHO (5-4) and t-HCOOH (5-4) show a similar distribution (Figure~\ref{fig:mom0}). Their integrated intensities are both brighter around the location of A1, and more compact than the CS (2-1) emission. Except for the CS emission towards A1, none of the tracers seem to peak at the location of A1 or A2. The moment 1 maps in Figure~\ref{fig:mom1} show that the ALMA observations presented here resolve the known velocity gradient towards source A
(\citealt{2012PinedaFirst,2016OyaInfalling,2020VantHoffTemperature}), approximately aligned with the major axis of the circumbinary disk-like structure. High-velocity blue-shifted and red-shifted components are clearly revealed in the CS (2-1) map towards A1 and A2, respectively. Higher-velocity components can also be identified in the other tracers, except for NH$_2$CHO, where the restricted velocity range used to avoid line contamination resulted in a removal of the high-velocity blue-shifted component (which is also present in this line).\\

Interestingly, the location where the integrated intensity of our molecular tracers is enhanced (particulary HNCO, NH$_2$CHO and t-HCOOH) coincides with the location of the dust substructures detected in the continuum emission (black contours in Figure~\ref{fig:mom0}). Similarly, outside the region in between the sources, the location of the higher-velocity material in the moment 1 maps coincides with the location of the dust substructures. The region covered by the dust substructures and the most compact molecular lines extends up to 20-40 au from the protostar positions. This is around the lower limit of the centrifugal barrier radius inferred for Source A in \cite{2016OyaInfalling}, from observations with a resolution of $\sim$ 70 au. In their model, the centrifugal barrier corresponds to a radius at which the infalling of material suddenly stops due to the conservation of angular momentum, within which a rotationally supported disk is expected to form. The infall velocity peaks at 
the centrifugal radius (twice the size of the centrifugal barrier). From there,  rotational motions dominate, with increasing velocity, down to the centrifugal barrier. Thus, higher velocities due to a combination of infall and rotation are expected in the region marking the transition from the inner envelope to the circumstellar disk. Although in a binary system the gas kinematics are expected to be more complex than in that simple model, accretion shocks and structures related to the transition from the circumbinary material to the circumstellar disks are also expected. In simulations, spiral structures connecting the circumbinary material to the circumstellar disks are observed \citep{2019MatsumotoStructure}. Further, \cite{2019MostaGas} find that these spiral structures can take the form of complex tightly wound features depending on the eccentricity of the system, the mass ratio and the specific orbital phase. That scenario has been given as a possible explanation for the curved filamentary features observed within the circumbinary material of the binary Class I system BHB2007-11 \citep{2019AlvesGas}, where the filamentary substructures are revealed at the same scales as the substructures in IRAS 16293 A. Thus the spatial correlation between the dust substructures and the enhancement of the integrated intensity and higher-velocities of our observed lines are consistent with these features tracing shocks, spiral or more complex features associated  to  the  transfer  of  material  from  the  circumbinary ambient into the circumstellar disks.\\

Figure~\ref{fig:mom1} also shows a clear almost 90$^{\circ}$ misalignment between the major axis of the circumstellar disk around A2 and the the direction of velocity gradient of the circumbinary material. Although such misalignment between the rotation axis of the circumbinary material and that of the region close to the protostars was expected given the known $\sim$ 90$^{\circ}$ difference in the P.A. of the bipolar outflows/ejection and the rotation axis towards source A down to 70 au scales \citep{1990MizunoRemarkable,2007LoinardNew,2014GirartOrigin,2012PinedaFirst,2016OyaInfalling,2019VanderWielALMAPILS}, the higher-resolution observations show that this misalignment persists down to the smallest scales resolved in our line observations ($\sim$13 au). This type of misalignment has been seen before for close protostellar binaries. For example, the Class I system IRS 43 with a separation of $\sim 74$ au \citep{2016BrinchMisaligned} in which the individual circumstellar disks were found to be significantly misaligned ($>60^{\circ}$), in inclination and P.A. Further, the orbital plane of the binary was constrained to be oriented close to face-on, while the circumbinary material was oriented close to edge-on. Misaligned configurations for the rotation axis of individual circumstellar disks, the circumbinary material and the orbital motion naturally arise in simulations where turbulence is included in the star forming cloud \citep{2010OffnerFormation,2018BateDiversity,2019LeeFormation}. For instance, members of a multiple system might form few thousands au apart, from gas with different angular momentum, and later move closer to form a bound tight binary (or higher multiplicity) system \citep{2016OffnerTurbulent,2018BateDiversity,2019KuffmeierBridge,2019LeeFormation}. Misalignment can also be the product of binary formation in an elongated structure whose minor axis is misaligned with the initial rotation axis \citep{1992BonnellFragmentation}. Finally, subsequent accretion of material with a misaligned angular momentum can also explain the misalignment between the compact dust disks and the surrounding rotating material, as observed towards IRAS 16293 A \citep{2018BateDiversity}.

\subsection{Line of sight velocities}
\label{sec:los}
\begin{figure*}[ht!]
\centering
\includegraphics[scale=0.8]{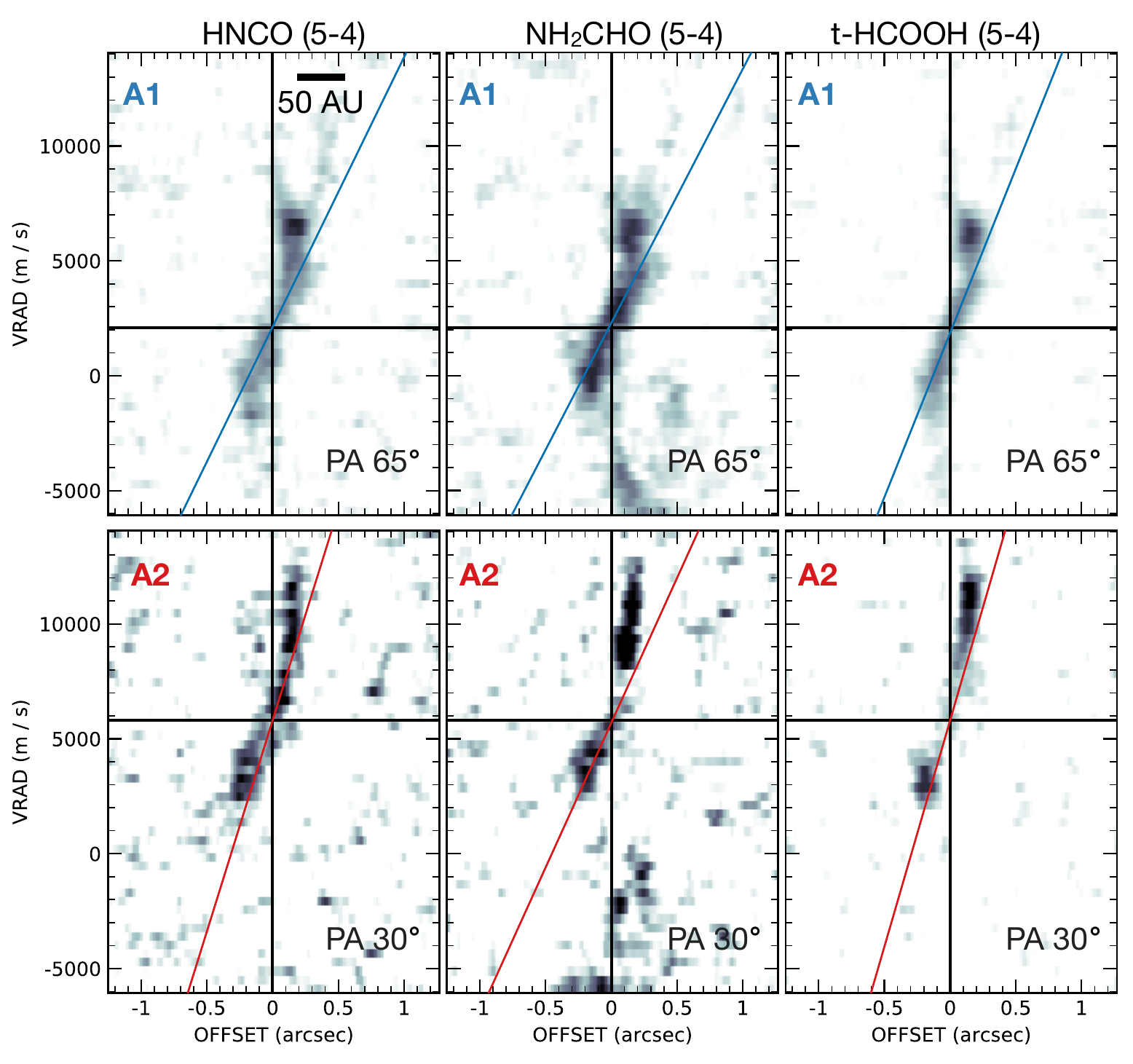}
\caption{Position-velocity diagrams of observed molecular lines centered around A1 (top) and A2 (bottom). The orientation of the cut for each source is indicated in the bottom right corner (see Appendix Figure~\ref{fig:cuts} for the orientation). The vertical black lines mark the position of A1 (top) and A2 (bottom), respectively. The inclined solid lines show the resultant linear gradient fit to each molecular line with which we obtain a line-of-sight velocity for each molecular line and source. The horizontal black lines show the average line-of-sight velocity of 2.1 \kms\ and 5.8 \kms\ calculated using the three molecular lines for A1 (top) and A2 (bottom). All the panels show the same velocity range and identical spatial scales. \label{fig:vsys_pvmaps}}
\end{figure*}

We use the HNCO (5-4), NH$_2$CHO (5-4) and t-HCOOH (5-4) lines which do not show significant artifacts from missing large scale structures to extract the velocity of A1 and A2 along the line-of-sight. To get a robust estimation of the sources velocities we use position-velocity (p-v) diagrams at a direction along the velocity gradient (Figure~\ref{fig:mom1}) around each source. The adopted P.A. corresponds to 65$^{\circ}$ and 30$^{\circ}$ for A1 and A2, respectively. The width of the p-v cuts was set to match the beam (5 pixels). See Figure~\ref{fig:cuts} for a diagram of the cuts overlaid on the moment 1 maps. The procedure to obtain the line-of-sight velocity consisted of fitting the p-v diagrams at intermediate velocities with a linear gradient and extract the velocity at the position of the source. For the linear gradient fit, in each channel along the p-v cut, we fit a Gaussian to the emission and then fit the peaks of the Gaussians. The velocity range that was fit with a linear gradient corresponds to 0.9-4 \kms\ and 4.7-8 \kms\ for A1 and A2, respectively. The selected range covers the region where the compact source is located in the p-v diagram as well as the region in which a single linear velocity gradient is observed. Figure~\ref{fig:vsys_pvmaps} shows the p-v diagrams for A1 and A2. The colored lines show the gradient fit for each molecular line.  The final line-of-sight velocity is given as the average and its associated error among the molecular lines for each source. Table~\ref{table:los} summarizes the results. We obtained line-of-sight systemic velocities of $2.1\pm0.1$ \kms\ and $5.8\pm0.1$ \kms\ for A1 and A2, respectively. We repeated the procedure using p-v diagrams with P.A. values differing from the previous one by $\pm$ 10$^{\circ}$. The resultant velocities are in agreement within uncertainties. \\

Our estimation of the line of sight velocity for A2 has the caveat that it uses the emission from tracers that were brighter around A1. If these tracers are not tracing closely the material near A2, the line of sight velocity might be different from the estimated value. Since the high-velocity red-shifted component approaching A2 in Figure~\ref{fig:mom1} starts at about 6 \kms, similar to the red-shifted CO emission from outflow lobes towards source A \citep{1990MizunoRemarkable}, the calculated line of sight velocity of $5.8$ \kms\ is consistent with an upper limit. On the other hand, \cite{2018DzibRevised} observed a water maser at a line of sight velocity of 2.1 \kms\ at a location consistent with A2 and moving along the direction of one of the CO molecular outflows. Thus we can consider the velocity of this water maser emission as a strict lower limit to the line of sight velocity for A2. This lower limit would result in both sources having the same line of sight velocity.

\begin{deluxetable}{lcccc}[ht]
\tablecaption{Line of sight velocities}             
\tablecolumns{5}
\label{table:los}      
\centering                          
\tablehead{\colhead{} & \colhead{HNCO (5-4)} & \colhead{NH$_2$CHO (5-4)} & \colhead{t-HCOOH (5-4)} & \colhead{Mean}\\
\colhead{} & \colhead{(\kms)} & \colhead{(\kms)} & \colhead{(\kms)} & \colhead{(\kms)}}
\startdata
A1&2.12 $\pm$ 0.07&2.30 $\pm$ 0.03&1.86 $\pm$ 0.04&2.09 $\pm$ 0.10\\
A2&5.90 $\pm$ 0.26&5.84 $\pm$ 0.21&5.70 $\pm$ 0.16&5.81 $\pm$ 0.05\\
\enddata
\end{deluxetable}

\subsection{Mass constraints from gas kinematics}
\label{sec:kep}

\begin{figure*}[ht!]
\centering
\includegraphics[scale=1]{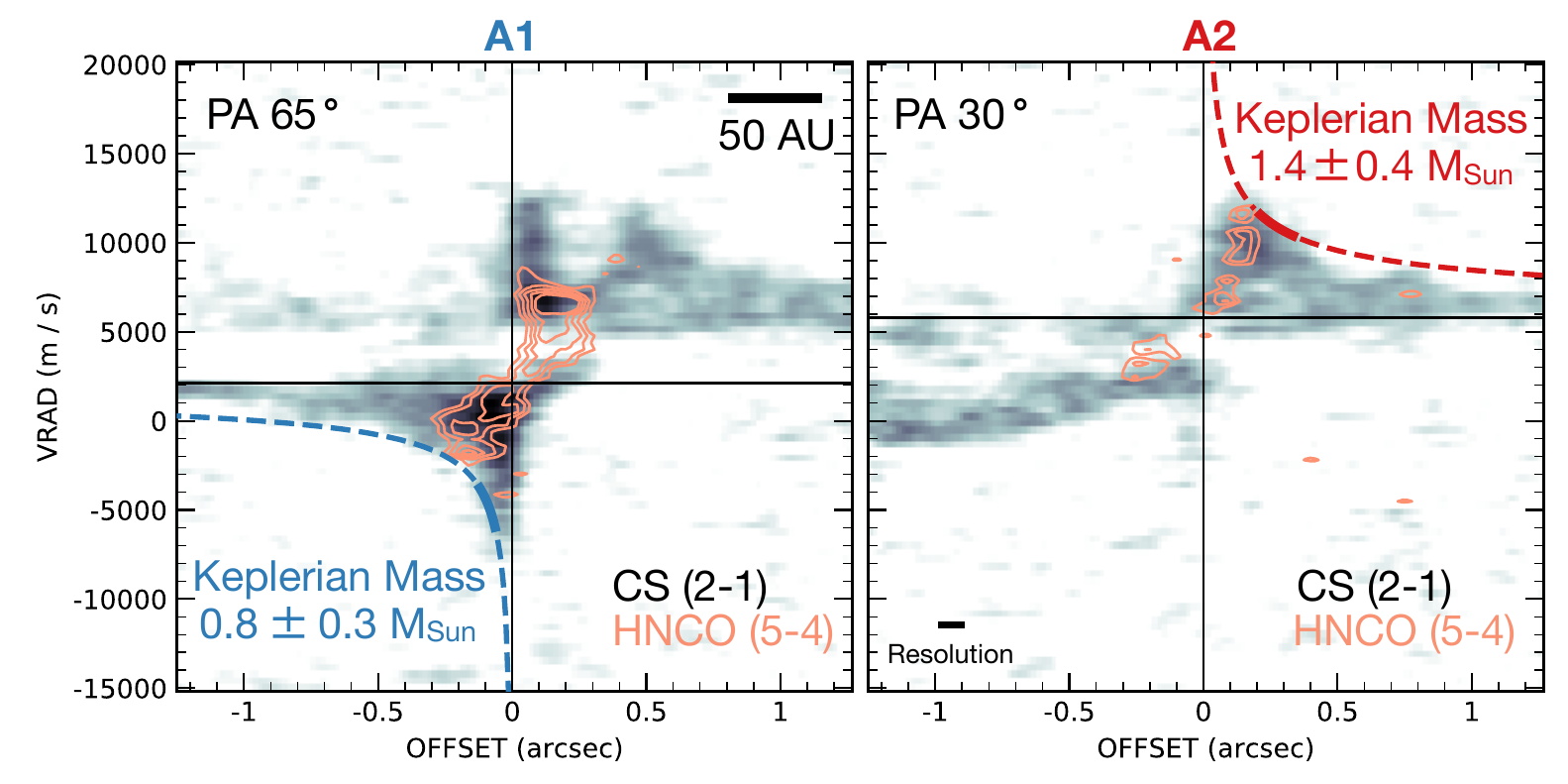}
\caption{Position-velocity diagrams of CS (2-1) centered around A1 (left) and A2 (right). The orientation of the cut for each source is in the top left corner and coincides with the orientation of the straight lines in Appendix Figure~\ref{fig:cuts}. The vertical black lines mark the position of the sources, while the horizontal black lines mark the systemic velocity for each source (Table~\ref{table:los}). The blue and red curves show the Keplerian rotation fit to the gas using only the bottom-left quadrant and upper-right quadrant for A1 and A2, respectively. The solid portion of the curve indicates the range of distances from the central source that were consistent with Keplerian rotation and hence used for the Keplerian profile fit. The dashed lines are an extrapolation of the solid curve Keplerian fit. The resultant masses are displayed in each case, corrected by inclination with $i=64^{\circ}$ (inclination of the extended emission surrounding the circumstellar disks, traced by the molecular lines and dust). Contours show the HNCO (5-4) counterpart emission (same as in Figure~\ref{fig:vsys_pvmaps}). In both panels the contours start at 3$\sigma$ and end at 7$\sigma$, increasing in steps of 1$\sigma=1.4$ mJy beam$^{-1}$. Both panels have the same velocity range and spatial scales.
 \label{fig:cs_pvmaps}}
\end{figure*}

To investigate the velocity profile of the material near A1 and A2 we use position-velocity diagrams along the gradient observed in the moment 1 maps. Although these gradients are misaligned with the major axis of the circumbinary disk-like material (P.A. $\sim$ 50$^{\circ}$) and the one of the circumstellar disk around A2 (P.A. $\sim$ 140$^{\circ}$), such misalignment does not rule out that the high-velocity components closest to the protostars might be tracing the Keplerian rotation of the gas circumstellar disks with misaligned axes (e.g., \citealt{2014JensenMisaligned}), a scenario that we test in the following sections.\\

Figure~\ref{fig:cs_pvmaps} shows CS (2-1) and HNCO (5-4) position velocity diagrams towards A1 and A2. The P.A. of the cuts are the same as those used for obtaining the line-of-sight velocities (see Figure~\ref{fig:vsys_pvmaps}) and correspond to direction along the velocity gradient at the location of each source (Appendix Figure~\ref{fig:cuts}). These directions are $15^{\circ}$ and $20^{\circ}$ different from the major axis of the circumbinary disk-like material, for A1 and A2, respectively. The HNCO (5-4) structures in the p-v maps are similar to that of the NH$_2$CHO (5-4) and t-HCOOH (5-4) (Figure~\ref{fig:vsys_pvmaps}). In this work, we focused on the gas kinematics close to the individual sources and located outside of the region in between the sources. This is because the region between the sources is likely more affected by the interaction between them. The CS (2-1) emission in the outer regions shows a resolved profile of increasing velocity towards A1 and A2 (in the bottom left quadrant for A1 and upper right quadrant for A2 in Figure~\ref{fig:cs_pvmaps}). In the following we analyze individually the velocity profiles observed in Figure~\ref{fig:cs_pvmaps} for A1 and A2. 

\subsubsection{A1}
\label{sec:a1_kine}
The CS (2-1) high-velocity blue-shifted component towards A1 can be identified down to a velocity of $\sim$ -8 \kms with the velocity peak located within 1 beam ($\lesssim$0.09" or 13 au) of the protostar position. On the other hand, the emission from the other lines represented in Figure~\ref{fig:cs_pvmaps} by HNCO (5-4), show a linear gradient around the protostellar location. The morphology of  HNCO (5-4) can be explained by emission arising from rotating gas concentrated at a narrow range of distances from the protostar (e.g., \citealt{2014LindbergALMA,2014YenALMAL1489IRS}). This is because if $v_{rot}$ is the rotation velocity of the rotating region at a radius $R$, then we expect the velocity across the major axis to change as $V_{rot}\times \Delta_{offset}/R$, with $\Delta_{offset}$ the distance to the source location. As discussed in Section~\ref{sec:moments}, the emission from HNCO might be tracing shocks or spiral features located at the transition from the circumbinary ambient into the circumstellar disks. The HNCO blue-shifted velocity peak is located farther away from the protostar than that of the CS. The latter could then be tracing the Keplerian motion of the circumstellar disk around A1. This is also in agreement with the interpretation of Keplerian motion of the H$_2$CS (7-6) in \cite{2016OyaInfalling}. This tracer showed similarly (although unresolved) high-velocity emission as CS (2-1) in this work. This high-velocity emission was inconsistent with the velocities of the farther away infalling and rotating material traced by other species.

Given the resolved velocity profile of the blue-shifted emission, we use the ‘upper edge’ method \citep{2016SeifriedRevealing} to extract the line-of-sight velocity of the gas as a function of the distance from A1. \cite{2016SeifriedRevealing} used simulations and showed that having a resolution of about 15 au is important to be able to identify a Keplerian profile using the 'upper' edge of the emission in a p-v diagram. Thus, our observations are adequate for applying this method to determine if the profiles are consistent with Keplerian rotation. We find that the velocity profile is indeed consistent with Keplerian rotation (i.e., $v \propto r^{-0.5}$) up to a distance of 20 au from A1. The extracted points beyond this distance depart from a Keplerian power law, in agreement with a steeper profile closer to $v \propto r^{-1}$. The data points consistent with a Keplerian power law and the correspondent fit are shown in Appendix Figure~\ref{fig:cs_upperedgefit}. The Keplerian profile fit assumes the systemic line-of-sight velocities in Table~\ref{table:los}. The resultant Keplerian curve is overlaid on the p-v diagram in Figure~\ref{fig:cs_pvmaps}. The solid part of the curve shows the regions used for the Keplerian fit, while the dashed lines are an extrapolation of the fit. From the Keplerian fit to the gas we obtain a mass of $0.8\pm0.04$ \msun for A1. This value is already corrected by the inferred inclination of the extended structure that the CS (2-1) line (and extended dust) is tracing (i.e., using $i=64^{\circ}$ ), as we do not resolve and/or detect line emission directly associated to the small dust circumstellar disk. We note that the inclination is only slightly different from the one inferred from the compact dust emission ($i=59^{\circ}$) and would change the resultant mass only by 10\%. As mentioned above, the method that we used to extract the velocity profile results in larger reported uncertainties on the masses (of a few tens of percent), when tested with simulations \cite{2016SeifriedRevealing}. Thus, we adopt here a larger $30\%$ uncertainty (based on Table 6 of Seifried et al. 2016), which results in an individual mass of $0.8\pm0.3$ \msun. By using a cut along the velocity gradient we minimize the contamination by infall motions, which nevertheless are likely present because the material is not oriented edge-on. We expect that the contamination by infall motions would change the mass within the large adopted uncertainty \citep{2016SeifriedRevealing}. As with the line-of-sight velocity analysis with p-v diagrams, we repeated the Keplerian fit procedure using p-v diagrams with P.A. values differing from the previous one by $\pm$ 10$^{\circ}$. The resultant masses were in agreement within the uncertainties of $30\%$. We note that although the $^{13}$CO (1-0) also shows a profile of increasing velocities towards A1 (Appendix Figure~\ref{fig:13co_mom}), the observations do not recover emission close to the source and thus, $^{13}$CO (1-0) does not trace the regions where the profiles were consistent with Keplerian rotation for CS (2-1).\\

Although we find that the velocity profile of the CS emission within 20 au is consistent with a Keplerian power-law, we cannot rule-out that this high-velocity material is tracing gas that is infalling from the edge of the circumbinary disk-like structure (traced by HNCO) due to, for example, a loss of angular momentum. In this case, the mass inferred by the Keplerian fit would be overestimating the mass. Using a simple infall and rotation model in Section~\ref{sec:kine_a2} for A2, we show that the overestimation factor is similar to the adopted 30\% uncertainty in the Keplerian mass. 

\subsubsection{A2}
\label{sec:kine_a2}
Figure~\ref{fig:cs_pvmaps} shows that the high-velocity red-shifted component can be identified up to a velocity of $\sim$13 \kms for CS, with the velocity peak located $\sim$0.14" or 20 au from the protostar. The emission from HNCO (5-4) is weaker near A2 compared with A1, but a linear gradient can also be identified (Figure~\ref{fig:vsys_pvmaps}). High-velocity red-shifted gas, reaching velocities similar to those of CS (2-1) are also observed in HNCO (5-4). We follow the procedure in Section~\ref{sec:a1_kine} to investigate if the profile could be consistent with a Keplerian power-law. In addition, since the velocity peak of CS, HNCO and the location of the dust substructure to the South-West of A2 coincide, the high-velocity peak in A2 might be associated with gas that is infalling and rotating from the circumbinary disk-like structure to the circumstellar disk. Thus, to estimate the mass in that scenario, we also use a simple model of infall and rotation (Appendix Section~\ref{ap:infall_rot}), similar to \cite{2016OyaInfalling}.\\

Following the procedure done for A1 (Section~\ref{sec:a1_kine}), we find that the velocity profile of CS towards A2 in Figure~\ref{fig:cs_pvmaps} is consistent with Keplerian rotation (i.e., $v \propto r^{-0.5}$) up to a distance of 50 au. The extracted points beyond these distances depart from a Keplerian power law, in agreement with a steeper profile closer to $v \propto r^{-1}$. The data points consistent with a Keplerian power law and the correspondent fit are shown in Appendix Figure~\ref{fig:cs_upperedgefit}. The resultant Keplerian curve is overlaid on the p-v diagram in Figure~\ref{fig:cs_pvmaps}. We obtain a mass of $1.4\pm0.4$ \msun. Similar to A1, we corrected by an inclination of $i=64^{\circ}$ and assumed an uncertainty of $30\%$ \citep{2016SeifriedRevealing}. We also repeated the Keplerian fit procedure using p-v diagrams with P.A. values differing from the previous one by $\pm$ 10$^{\circ}$. The resultant masses were in agreement within the uncertainties of $30\%$. For this source, the $^{13}$CO (1-0) (Appendix Figure~\ref{fig:13co_mom}) shows a velocity profile similar to that of CS (2-1).\\

\begin{figure}[ht!]
\centering
\includegraphics[scale=0.9]{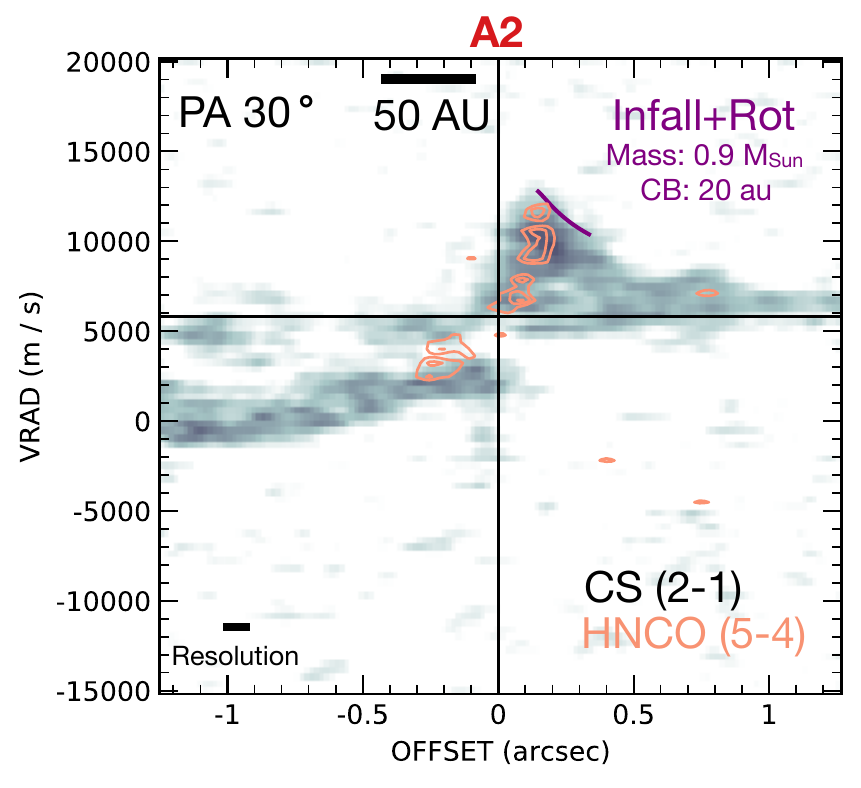}
\caption{Position-velocity diagrams of CS (2-1) and HNCO (5-4) centered around A2, same as right panel in Figure~\ref{fig:cs_pvmaps}. The purple curve shows the velocity obtained considering infall with rotation with conservation of angular momentum around a central mass of 0.9 \msun, and a centrifugal barrier of 20 au (where the innermost part of the purple profile ends). The curved is corrected by inclination and projection effects considering that the rotation axis is perpendicular to the major axis of the circumbinary material (P.A. $=50^{\circ}$). The kinematics of an infalling and rotating flow of gas can also (in addition to pure Keplerian motion) explain the velocity profile towards A2. In both cases the central mass agree within the uncertainties.\label{fig:infallrot_A2}}
\end{figure}

For the combination of rotation and infall scenario, we compared the velocity profile with a curve of a simple model of rotating material undergoing infall with conservation of angular momentum described in detail in Appendix Section~\ref{ap:infall_rot}. This simple model is similar to the one used to provide the size of a centrifugal barrier for IRAS 16293 A by \cite{2016OyaInfalling}, as well as to describe in general the kinematics outside of the Keplerian rotation region in other protostellar sources \citep{2014YenALMAL1489IRS,2014NaturSakai}. By assuming that the rotation axis of the material is perpendicular to the major axis of the circumbinary disk-like structure, we find that a centrifugal barrier of 20 au and mass of 0.9 $\msun$ can reasonably reproduce the velocity profile along the same cut used for the Keplerian fit (see Figure~\ref{fig:infallrot_A2}). This profile assumes the same inclination as the one used for the Keplerian fit ($i=64^{\circ}$). A smaller centrifugal barrier can also reproduce the profile if the central mass is increased, while a larger centrifugal barrier cannot reproduce emission that is as close to the protostar as the one observed in Figure~\ref{fig:infallrot_A2}. We note that the value of the masses inferred in the two cases agree under the adopted uncertainty of few tens of percent.

\section{Discussion}
\label{sec:discussion}
An important conclusion from our kinematics analysis is that the point sources masses are consistently larger than previous estimations with 5-6 times lower resolution, in which the location of the point sources were not resolved \cite{2012PinedaFirst,2016OyaInfalling,2018JacobsenALMAPILS}. They derived a mass of $0.8$ \msun, assuming a single source. This is in agreement with our results within the uncertainties. However, recent work comparing observations and synthetic observations of the kinematics of the circumbinary and circumstellar disks for the binary system L1551 NE further support our finding of higher protostellar mass for IRAS 16293 A than previously estimated. L1551 NE is a Class I system at a similar distance, showing a similar separation and inclination as IRAS 16293 A. \cite{2017TakakuwaSpiral} find maximum molecular line velocity differences (with respect to the protostellar system velocity) of about 4 \kms, while the protostars line of sight velocity difference is 1.3 \kms, compared to about 8 \kms and 3.7 \kms, respectively, for IRAS 16293 A. \cite{2017TakakuwaSpiral} find that a simulation with a total protostellar mass of 0.8 \msun is in agreement with the observed kinematics. By comparison, the higher velocity differences observed towards IRAS 16293 A would imply a higher mass, consistent with our results. Future comparisons with synthetic observations as in \cite{2017TakakuwaSpiral} will help to further constrain the individual masses.\\

\begin{figure*}[ht!]
\centering
\includegraphics[width=\textwidth]{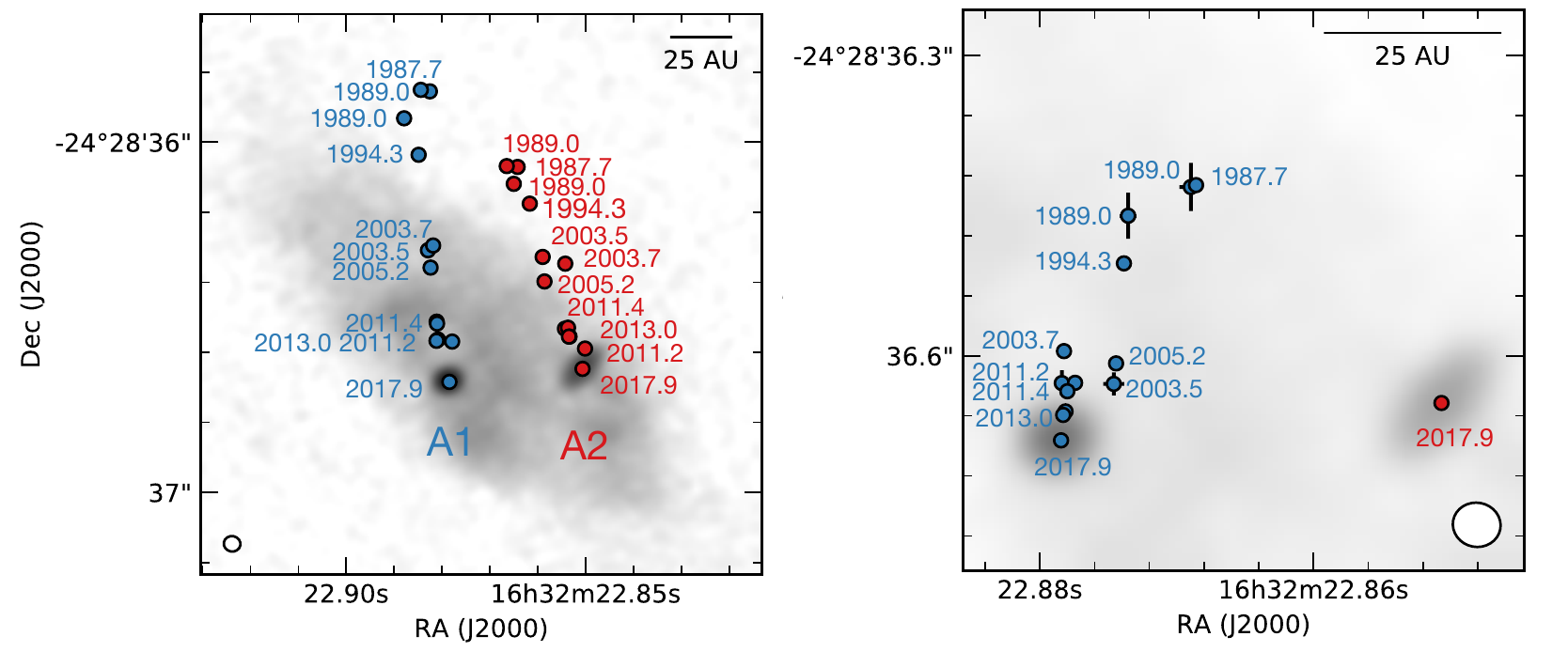}
\caption{Motions of A1 and A2 over a 30 year period. All observations prior to 2017 correspond to VLA observations recently compiled in \cite{2019HernandezGomezNature}, and in which by visual inspection the two sources were resolved and free from ejecta contamination. Left: Absolute motions of A1 and A2 overlaid on the ALMA 3 mm observation. Right: Relative motions of A1 with respect to A2 overlaid on the ALMA 3 mm observation. When errors are not visible, it is because they are smaller than the symbol.
\label{fig:positions_abs_rel}}
\end{figure*}

\subsection{Are A1 and A2 bound?}
\label{sec:bound_analysis}
For observed embedded multiple systems with separations $<$ 100 au the usual assumption is that they are bound \citep{2016TobinVLA}. For this close binary pair, we have measured their motions and thus we can estimate the required total mass for the pair to be bound. We use the most-recent published proper motions for A1 and A2 \citep{2019HernandezGomezNature} that used data spanning almost 30 years, along with our new line-of-sight velocity to obtain the total velocity for each source. The measured individual line-of-sight velocities in Section~\ref{sec:los} lead to a relative line of sight difference between A1 and A2 of 3.7 \kms. The relative velocity (A1-A2) on the line-of-sight and plane-of-sky is -3.7 $\pm$ 0.2 \kms\ and 5.2 $\pm$ 0.6 \kms, respectively, yielding a total relative velocity magnitude of 6.4 $\pm$ 0.5 \kms. The minimum mass for A1 and A2 to be a bound pair follows from the following condition:
\begin{equation}
    E_{kin}+E_{pot}\leqslant0
\end{equation}
where the kinetic and potential energy are given by:
\begin{eqnarray}
E_{kin}&=&\frac{1}{2}\sum_{1,2}M_i(\vec{v_i}-\vec{v_{com}})^2 \\ 
E_{pot}&=&\frac{-GM_1M_2}{|\vec{r_1}-\vec{r_2}|}
\end{eqnarray}

$M_i$ corresponds to the mass of source $i$, $v_i$ is the velocity of source $i$, $v_{com}$ is the velocity of the center of mass and $|\vec{r_1}-\vec{r_2}|$ is the separation between the sources. The minimum total mass can be thus calculated as:
\begin{equation}
\frac{1}{2}\frac{(\vec{v_1}-\vec{v_2})^2}{G/|\vec{r_1}-\vec{r_2}|}\leqslant M_1+M_2
\end{equation}

Assuming that the point sources lie in the plane of the circumbinary disk-like material for which we calculated a P.A. of $50^{\circ}$ and inclination with respect to the plane-of-sky of $64^{\circ}$ (Section~\ref{sec:cont}), the deprojected distance between A1 and A2 is 0.678" or 95.6 au. This results in a minimum mass of 2.2 $\pm$ 0.3 \msun. As discuss in Section~\ref{sec:los}, the line of sight velocity difference between A1 and A2 might be smaller. If we consider the line of sight velocity difference to be zero we obtain a strict minimum mass of 1.7$\pm$0.3 \msun\ for the system to be bound. In Section~\ref{sec:kep} we used the gas kinematics and estimated a combined A1+A2 mass of $2.2\pm0.5$, suggesting that the pair is indeed bound. 

\subsection{Motion of the protostars}
\label{sec:stellar_kine}

\begin{figure*}
\centering
\includegraphics[scale=1]{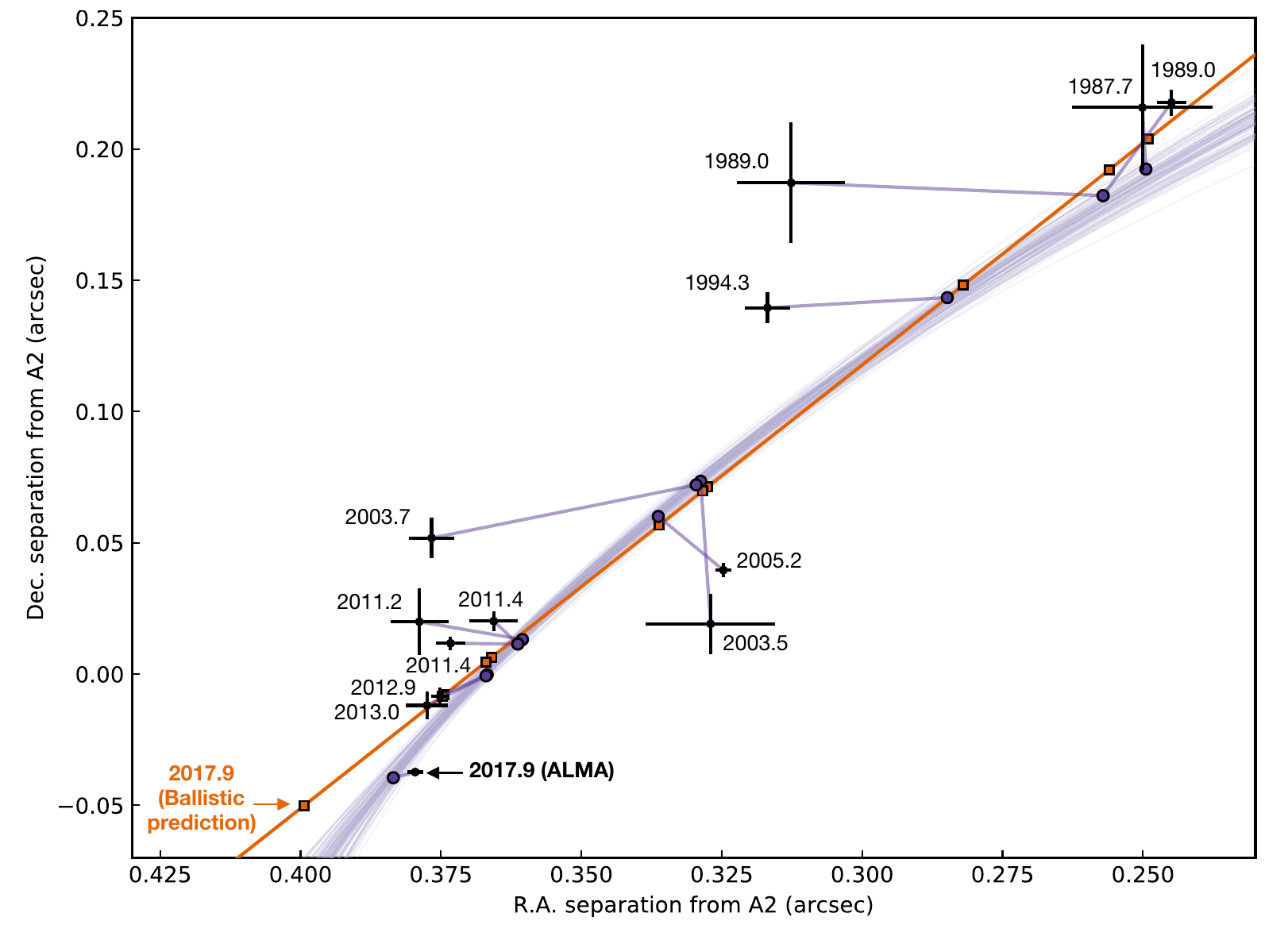}
\caption{Relative motions of A1 with respect to A2 (black markers with errors). The straight line shows the best ballistic trajectory using only the VLA data. The square symbols overlaid on the straight line show the predicted positions for all observations along the ballistic trajectory. The purple curves show one hundred bound Keplerian orbital solutions drawn from posterior distributions (Appendix Figure~\ref{fig:corner}). The dot symbols along the orbital trajectory show the predicted positions for all observed positions along a randomly selected Keplerian orbital trajectory. Lines connecting the observed positions with the predicted ones along the randomly selected Keplerian orbit are also shown for clarity. When errors are not visible, it is because they are smaller than the symbol. \label{fig:orbits}}
\end{figure*}

Further insight into the dynamics of this now bound binary system can be obtained through the study of their proper motion. We examined all VLA positions preceding the ALMA measurement and reported in \cite{2019HernandezGomezNature}. We selected from this work (and references therein) only those observations for which by eye the two sources were resolved and free from ejecta contamination. These criteria resulted in twelve VLA observations with the epochs listed in Appendix Table~\ref{tb:vla_data}. Ejecta emission from A2 is routinely seen in VLA 22 GHz observations since 2006 \citep{2007LoinardNew}. As a result of this, all selected observations after 2005 correspond to frequencies $>$33 GHz. The left panel in Figure~\ref{fig:positions_abs_rel} shows our selected absolute positions for A1 and A2 through time, overlaid on the most recent ALMA 3 mm continuum observation. A1 has moved faster in the plane of the sky than A2, resulting not only in a change in orientation but also in the separation between the sources. The separation has changed from 47 au in the first observation to 54 au in the most recent one. These changes are clearly seen in the right panel in Figure~\ref{fig:positions_abs_rel} showing the relative position of A1 with respect to A2 through time.
We note that the relative positions are not affected by differences in the absolute astrometry accuracy of the observations. The change in time of the relative positions is important since if A1 and A2 are orbiting around their center of mass, the relative trajectory of A1 with respect to A2 is also described by a Keplerian orbit which parameters also provide the total mass of the system, including contributions from both protostars+compact disks, and the gas mass enclosed within the orbit in the case of this young embedded system. The relative motions in Figure~\ref{fig:positions_abs_rel} further suggest they are tracing a bound orbital trajectory. In agreement with this expectation, \cite{2019HernandezGomezNature} tried a quadratic fit to the relative A1-A2 positions. Although that parabolic trajectory provided a better fit to their data (particularly their most recent data), our ALMA observation is in disagreement with the prediction of their quadratic fit. The quadratic fit predicts a reversal in the declination motion of A1 with respect to A2 around 2015, indicating that A1 started to move towards the North (see their Figure 7), while our new ALMA data point shows that A1 kept moving South (Figure~\ref{fig:positions_abs_rel}). Such discrepancy might be due to the use of jet contaminated and/or low resolution data which strongly overwhelms the subtle astrometric variations due to the orbital motions and that were discarded in our analysis. Despite our careful selection, we note that the errors in Figure~\ref{fig:positions_abs_rel} are only those from the Gaussian fit to the compact sources and thus do not take into account errors produced by unresolved ejecta emission that can temporarily displace the observed center from the true center. As data earlier than 2010 had typically lower resolution (Table~\ref{tb:vla_data}), it is likely that the differences between positions at closely separated epochs or among different epochs in the early VLA data (e.g., epoch 1989 in Figure~\ref{fig:positions_abs_rel}) are due to this temporary “wobble” from unresolved ejecta.\\

To further investigate the relative A1-A2 trajectory in light of the new ALMA observations we first calculated the best ballistic trajectory using our selected VLA epochs. Then, we compared the predicted position along this trajectory with the observed ALMA position. Figure~\ref{fig:orbits} shows the relative positions and the calculated ballistic trajectory (straight line). The new ALMA epoch is in disagreement with the prediction from the ballistic trajectory. Given this and our previous conclusion that the system is likely bound (Section~\ref{sec:bound_analysis}), we also fit all relative positions (VLA+ALMA) with Keplerian orbital trajectories using a Monte Carlo approach (see Appendix Section~\ref{sec:orb_method} for details of the fitting procedure and posterior distributions). Figure~\ref{fig:orbits} shows the most likely Keplerian orbital trajectories along with the ballistic line. The predictions for all epochs along both trajectories (ballistic and Keplerian orbit) are also marked. Both trajectories have similar predicted positions before the ALMA observation. If we re-calculate the best ballistic trajectory using all epochs we obtain sightly better residuals for the Keplerian orbital trajectory. Yet, the Keplerian orbital trajectory has seven free parameters, while the ballistic trajectory only four. Thus, although the current data do not allow us to securely rule-out the ballistic trajectory, the results from the gas kinematics provide independent evidence in support of the Keplerian trajectory. Future monitoring observations are key to further strengthen this conclusion while also allowing improved accuracy on the Keplerian orbit determination. 


The full posterior distributions, medians and confidence intervals of the orbital parameters are given in Appendix Section~\ref{sec:orb_method} and displayed in Figure~\ref{fig:corner}. The resultant orbits have a period of $P=362^{+133}_{-73}$ years and semimajor axis of $a=80.26^{+14.60}_{-8.81}$ au. The inclination also seems to be well constrained ($i=58.69\degr ^{+3.39}_{-3.82}$) and is similar to the inclination of the circumbinary gas and dust emission. The other angles which are required to fully define the orientation of the orbit need further observations to have accurate constraints. Similar is the case of the derived eccentricity ($e=0.19^{+0.09}_{-0.06}$). This is because tests in which a couple of epochs were not considered (2005.2 and 2003.5) provided different values for these parameters, while the rest remained consistent. The total mass derived from the Keplerian orbital trajectory is $M_{tot}=4^{+1}_{-1}$ \msun. This derived total mass from the orbit is in agreement with previous estimations assuming a simplified plane of the sky circular orbit (\citealt{2005ChandlerIRAS16293,2007LoinardNew,2010PechConfirmation}), although in these studies the protostellar nature of A1 had not been confirmed. Since the presence of gas is also contributing to the total derived mass, we can take this value along with its large uncertainty as an upper limit to the combined mass of the point sources. In addition, we can also provide an independent upper limit using the luminosity of Source A. This source has been modeled by \cite{2018JacobsenALMAPILS}, although only as a single source, and they found that a luminosity of 18 L$_{\odot}$ resulted in good agreement with their observations. Since a 2 \msun\ pre-main sequence star has an approximate luminosity of $\sim$10 L$_{\odot}$ at the birthline \citep{2005StahlerFormation}, we can set this value as the upper limit to the most massive protostar. Thus the combined evidence, from the gas kinematics, stellar kinematics, and luminosity results in individual protostellar masses reasonably constrained in the range $0.5\lesssim M_1\lesssim M_2 \lesssim2$ \msun. The order of the previous relation takes into account that relative to Source B,  A1 has moved significantly more than A2 (\citealt{2002LoinardLarge,2005ChandlerIRAS16293,2010PechConfirmation}). These constraints result in a mass ratio between $\sim$ 0.3 and 1.\\

Further constraints on the mass ratio could be obtained by also deriving the Keplerian orbital parameters of the center of mass of Source A with respect to Source B which are also consistent with being bound (see Appendix Section~\ref{sec:orb_method}), forming a triple hierarchical system. However, the separation between A and B is an order of magnitude larger than the separation between A1 and A2, resulting in possible orbits between the center of mass of A relative to B of several $10^3-10^4$ years. Thus we cannot constrain this orbit with the current 30 years of observations. 

\section{Summary and Conclusions}
\label{sec:sum_conclusions}
We use high resolution continuum (6.5 au resolution) and line (13 au resolution) 3 mm ALMA observations towards the Class 0 multiple system IRAS 16293-2422. In this work, we analyzed the southern source in this system (IRAS 16293 A). In addition, we use observations from the VLA covering a period of 30 years \citep{1989WoottenDuplicity,2019HernandezGomezNature} to review the motion of the compact sources within IRAS 16293 A. Our results can be summarized as follows:

\begin{itemize}
\item The two radio sources A1 and A2 are unambiguously detected in the 3 mm continuum observations and a considerable fraction of the flux in both sources is consistent with thermal dust emission. Thus, the 3 mm observations confirm the protostellar nature of both sources, which remained debated due to conflicting results between previous $>200$ GHz and $<43 $ GHz VLA observations. The peaks (Aa, Aa* and Ab) observed at $>200$ GHz were likely affected by optical depth which prevented the clear detection of the embedded compact sources. Some of these peaks are tracing substructures in the extended emission instead.

\item The compact emission towards A2 is resolved and is consistent with a dust circumstellar disk with a FWHM size of 12 au, oriented perpendicular to the previously observed bipolar ejecta at cm wavelengths as well as perpendicular to the disk-like extended circumbinary dust emission. The compact emission towards A1 is unresolved, setting a limit to the FWHM size of the dust circumstellar disk of 3.6 au. 

\item Complex substructures extending from 20-40 au from the protostars are also observed. They are associated to regions where the emission of several lines of the $J=5-4$ rotational transition of HNCO, NH$_2$CHO and t-HCOOH is enhanced. Similarly, they are associated with regions where these tracers as well as CS (2-1), show higher velocities. Thus, these substructures might be tracing shocks or spiral features at the transition from the circumbinary structure into the circumstellar disks.

\item We use the compact emission from HNCO (5-4), NH$_2$CHO (5-4) and t-HCOOH (5-4) to estimate individual line-of-sight velocities for A1 and A2 yielding a line-of-sight velocity difference of 3.7 \kms. CS (2-1) traces clear high-velocity emission associated with the positions of the protostars. The velocity profiles from the locations of the sources and up to few tens of au towards the outskirts are consistent with a Keplerian power-law. The Keplerian power-law implies individual masses of $0.8 \pm 0.3$ \msun\ and $1.4 \pm 0.4$ \msun\ for A1 and A2, respectively. The velocity profiles can also be explained by material that is rotating and infalling from the circumbinary disk-like structure to the circumstellar disks resulting in smaller although comparable masses.

\item The most recent reported proper motions from VLA observations, our newly measured line-of-sight velocities and protostellar masses from the gas kinematics indicate that the binary system A1-A2 is bound. 

\item The new positions from the ALMA observations depart from the predicted position along a ballistic trajectory inferred from the VLA observations, suggesting the observation of an orbital trajectory. We fit orbital parameters to the relative positions of the VLA+ALMA observations resulting in orbital solutions with a period of $362^{+133}_{-73}$ years, semi-major axis of $80.26^{+14.60}_{-8.81}$ au and inclination consistent with that of the extended circumbinary material. The results also indicate a low eccentricity ($e=0.19^{+0.09}_{-0.06}$) but future observations are needed to better constrain the geometry of the orbit. The total mass derived from the orbital fit is $M_{tot}=4^{+1}_{-1}$ \msun. The independent mass constraints from the gas kinematics and the stellar kinematics are in agreement within the uncertainties, which when added to luminosity restrictions results in individual masses reasonably constrained in the range $0.5\lesssim M_1\lesssim M_2 \lesssim2$ \msun.\\
    
\end{itemize}

The range of protostellar masses inferred from the orbital analysis and the gas kinematics are consistently higher than previous estimations using lower resolution observations of the gas kinematics or models with a single source. Given the current mass of the IRAS 16293 A and B envelope of $5 \pm 1$ \msun at scales of a few 1,000 au, the binary system A and single source B are also likely bound, forming a triple hierarchical system. Future monitoring observations, as well as detailed modeling with simulations, will help to further constrain the dynamics and individual masses of this deeply embedded triple system. 



\acknowledgments
We thank the anonymous referee for the constructive comments that helped improve the overall content of the manuscript. The authors thank Bo Zhao and Kedron Silsbee for insightful theoretical discussions on binary formation and orbits. LL acknowledges the financial support of DGAPA, UNAM (project IN112417), and CONACyT, M\'exico.  This project has received funding from the European Union's Horizon 2020 research and innovation programme under the Marie Skłodowska-Curie grant agreement N$^{\circ}$ 823823 (DUSTBUSTERS).\\

This paper makes use of the following ALMA data: ADS/JAO.ALMA\#2017.1.01247.S. ALMA is a partnership of ESO (representing its member states), NSF (USA) and NINS (Japan), together with NRC (Canada), MOST and ASIAA (Taiwan), and KASI (Republic of Korea), in cooperation with the Republic of Chile. The Joint ALMA Observatory is operated by ESO, AUI/NRAO and NAOJ.

%







\appendix
\renewcommand{\thefigure}{A\arabic{figure}}
\renewcommand{\theHfigure}{A\arabic{figure}}
\setcounter{figure}{0}

\section{Molecular line properties}

Table~\ref{tb:lines} summarizes the frequencies and upper energy levels of the transitions analyzed in this work. This list does not include all the lines in the data, as there are several weaker lines, which were not suitable for a kinematical analysis and whose identification was beyond the scope of this study.\\

\begin{deluxetable}{lccc}[h!]
\tablecaption{Properties of the observed transitions}             
\label{tb:lines}      
\centering                          
\tablecolumns{4}
\tablewidth{0pt}
\tablehead{
\colhead{Molecule} & \colhead{Transition} & \colhead{Frequency (GHz)} & \colhead{E$_{up}$ (K)}
}
\startdata
CS & 2-1 & 97.980953 & 7\\
HNCO & 5$_{3, 2}$-4$_{3,1}$ & 109.833487 & 391\\
HNCO & 5$_{3, 3}$-4$_{3,3}$ & 109.833487 & 391\\
NH$_2$CHO & 5$_{1, 4}$-4$_{1,3}$ & 109.753578 & 19\\
t-HCOOH & 5$_{2, 4}$-4$_{2,3}$ & 112.287144& 29\\
t-HCOOH & 5$_{4, 1}$-4$_{4,0}$ & 112.432319\tablenotemark{a} & 67\\ 
t-HCOOH & 5$_{3, 3}$-4$_{3,2}$ & 112.459621 & 45\\
t-HCOOH & 5$_{3, 2}$-4$_{3,1}$ & 112.467007 & 45\\
$^{13}$CO & 1-0 & 110.201354 & 5\\
C$^{17}$O\tablenotemark{b} & 1-0 & 112.358982 & 5\\
C$^{18}$O\tablenotemark{c} & 1-0 & 109.782173 & 5\\
\enddata
\tablecomments{All the cubes were imaged with a channel width of 0.38 \kms. Frequencies and transitions were based on the full CDMS and JPL catalogues available within the CASSIS software (developed by IRAP-UPS/CNRS \url{http://cassis.irap.omp.eu}).}
\tablenotetext{a}{There is an additional transition (t-HCOOH 5$_{4, 2}$-4$_{4,1}$) separated by 0.07 \kms, and thus unresolved in our observations.}
\tablenotetext{b}{Undetected.}
\tablenotetext{c}{Detected, emission is very weak and extended and thus was not used in this study.}
\end{deluxetable}



\renewcommand{\thefigure}{B\arabic{figure}}
\renewcommand{\theHfigure}{B\arabic{figure}}
\setcounter{figure}{0}
\section{Fits to the compact emission towards A1 and A2}
\label{ap:contfit}
\begin{figure*}[h!]
   \centering
   \includegraphics[scale=0.65]{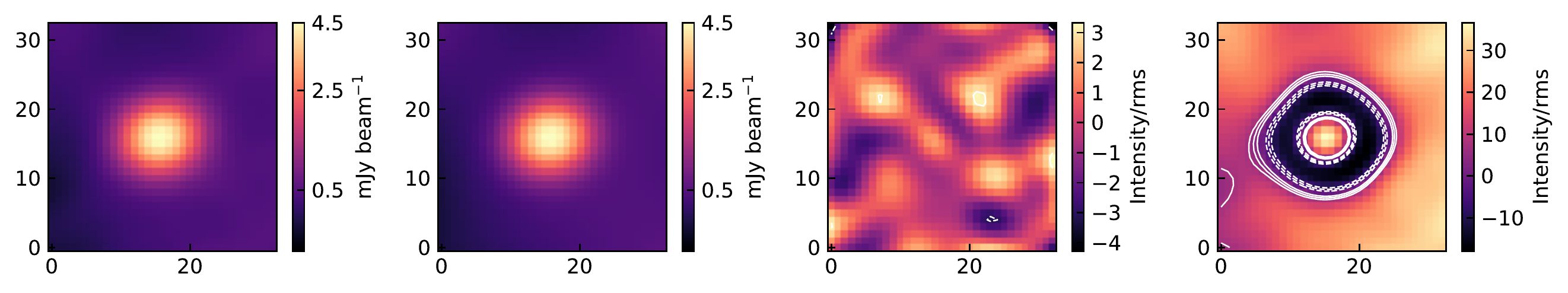}
   \includegraphics[scale=0.65]{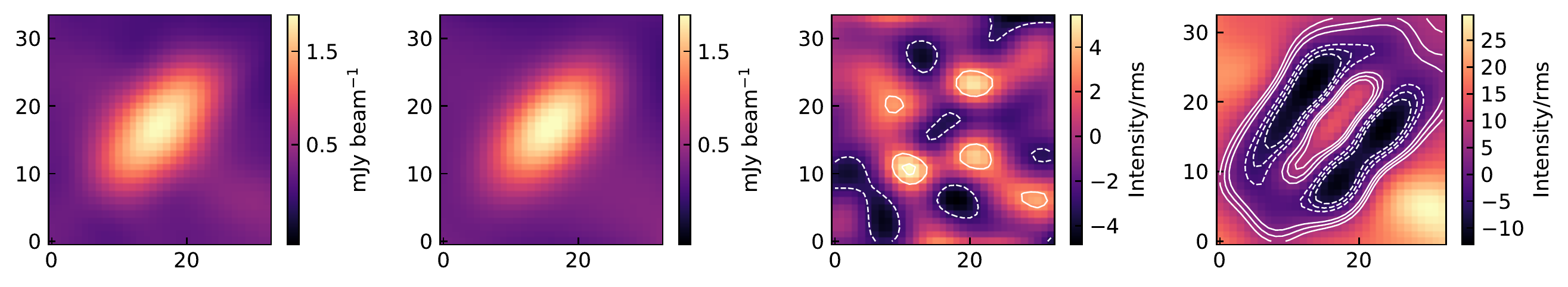}
   \caption{Observed 3 mm compact emission and comparison of residuals from the Gaussian fit with and without background subtraction for A1 (top) and A2 (bottom). From left to right the panels show the observed 3 mm emission, the Gaussian fit to the compact emission plus the background fit, the residuals between the observed emission and the Gaussian plus background fit and the residuals when doing the Gaussian fit without background subtraction. In all panels, x-axis and y-axis are in units of pixels. Contours in the fourth and fifth columns correspond to sigma levels of $-7,-5,-3, 3, 5, 7$. Negative sigma levels are shown as dashed contours.}
   \label{fig:cont_maps_fit}
    \end{figure*}

We carried out 2D Gaussian fits in the image plane of the bright and compact continuum emission towards A1 and A2. We performed fits with and without background subtraction. For the fit without any background subtraction, we use the CASA task imfit, restricting the fitting region to a square of size 0.18". For the fit with background subtraction, we extracted a square sub-image around each source with a size of 0.20". We then made a spline fit of the background with the scipy task bisplrep, masking out the compact emission. The mask for source A1 consisted of a circular region of radius 0.07", while for A2 consisted of an ellipse following the orientation of the compact emission with semimajor and semi-minor axes of 0.11" and 0.07”, respectively. We checked that the background fit shows no hints of compact emission at the center. We then subtracted the background fit from the data and performed a 2D Gaussian fit with imfit without any restriction. Figure~\ref{fig:cont_maps_fit} shows the data, the fit with background subtraction and the residuals of the fit with and without background subtraction for both sources. The position of the peak and the peak flux do not change with the type of fit while the size and integrated flux are 1.5-2$\times$ smaller when the fit is done after background subtraction. Since the residuals for the fit with background subtraction are substantially better, we list the results of the background subtraction fit in Table~\ref{tb:3mmfit} and we use these fluxes for the mass estimation. The positional errors are $\sim1\times10^{-4}$ arcseconds and $\sim7\times10^{-4}$ arcseconds for A1 and A2, respectively. \\

For assessing the free-free contribution, we also attempted a two components fit (Gaussian+point source) to the compact 3 mm emission towards A1 and A2, after background subtraction. For A1, all the properties of the Gaussian except for the peak flux and integrated flux remain consistent with the previous single Gaussian fit. The peak and integrated flux are reduced in a 25\% and 30\%, respectively. This is similar to the free-free contamination ($\sim$37\%) estimated using the A1 radio spectral index obtained from VLA observations \citep{2019HernandezGomezNature}. For A2, the resultant peak and integrated fluxes for the point source were consistent with zero, while the free-free contamination using the A2 radio spectral index \citep{2019HernandezGomezNature} was $\sim$53\%.

\begin{figure}[ht!]
\includegraphics[width=0.4\textwidth]{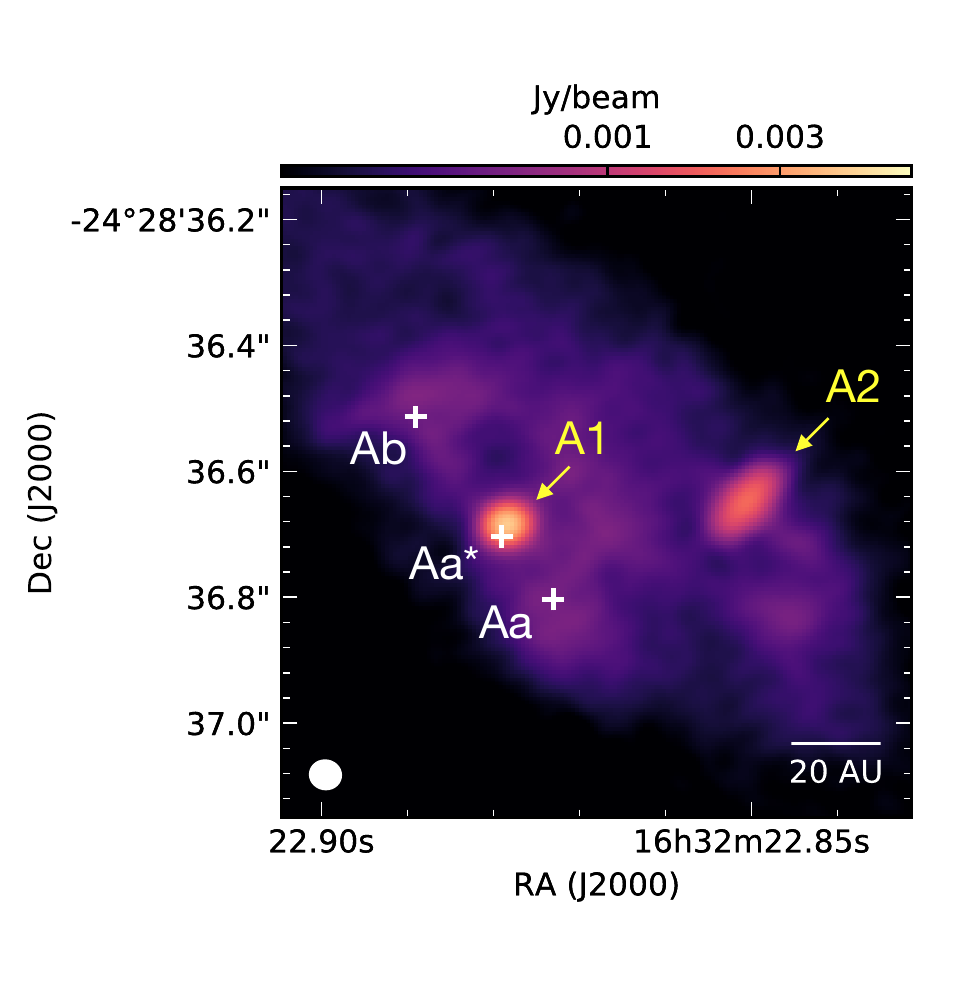}
\caption{Zoom-in view towards source A in the ALMA 100 GHz observations. The bright and compact emission correspond to the protostellar sources A1 and A2 (epoch 2017.9). The crosses mark the positions of the previously detected peaks using lower resolution ($\gtrsim$ 18 au) observations at frequencies $>200$ GHz \citep{2005ChandlerIRAS16293,2013ChenSMA,2018SadavoyDustIRAS16293}. The emission at frequencies $>200$ GHz is optically thick and does not let one see the peaks tracing the compact sources embedded (particularly A2), which are clearly revealed in the 100 GHz observations, even when degrading the resolution to match previous observations. The positions for Aa, Ab, and Aa* correspond to those reported in \cite{2018SadavoyDustIRAS16293} (epoch 2015.5) recalculated with respect to the peak of source B to account for proper motion between the epochs (although small$\sim$0.01"). While Ab and Aa are tracing the substructures on the extended disk-like structure, Aa* (the weakest among the three) peaks near A1. However, Aa* peak location (to the South of A1 in the figure) is inconsistent with the trajectories discussed in Section~\ref{sec:stellar_kine} and the motions in Figure~\ref{fig:positions_abs_rel}, and thus not a reliable tracer of A1. \label{fig:chand_peaks}}
\end{figure}

\section{Simple velocity profile of infalling and rotating gas}
\label{ap:infall_rot}
We use the simple model in which the gas motion is approximated by particles
moving in a plane under the influence of a central object of mass $M$. The particles are considered to have a constant angular momentum (perpendicular to the plane of motion) along the trajectories, resulting in a centrifugal barrier of radius:

\begin{equation}
    r_{CB}=\frac{j^2}{2GM},
\end{equation}

where $j$ is the specific angular momentum. This is equivalent to the trajectories calculated by \cite{1976UlrichInfall}, if we consider only the motion in the plane perpendicular to the rotation axis (equatorial plane). The infall $v_{inf}$ and rotational  $v_{rot}$ velocities are given by:

\begin{equation}
    v_{inf}=\sqrt{\frac{2GM}{r}-\frac{j^2}{r^2}} ,
\end{equation}
\begin{equation}
    v_{rot}=\frac{j}{r} ,
\end{equation}

where $r$ is the distance to the protostar calculated as

\begin{equation}
    r=\sqrt{x^2+y^2}
\end{equation}
\begin{equation}
    x=d\times\Delta_{off}\cos{(P.A.-P.A.^{\prime})}
\end{equation}
\begin{equation}
    y=d\times\Delta_{off}\sin{(P.A.-P.A.^{\prime})}/\cos{(i)} ,
\end{equation}
with $d$ the distance to the source, $\Delta_{off}$ the offset in arcseconds along the position-velocity cut with a position angle $P.A.$, and $i$ the inclination with respect to the plane-of-sky. $P.A.^{\prime}$ corresponds to the position angle of the major axis of the equatorial plane. Then, the final velocity profile $v_{los}$ on the p-v map will be given by:

\begin{equation}
    v_{los}=v_{inf}\frac{y}{r}+v_{rot}\frac{x}{r}
\end{equation}

\renewcommand{\thefigure}{D\arabic{figure}}
\renewcommand{\theHfigure}{D\arabic{figure}}
\setcounter{figure}{0}
\section{Orbit fitting}
\label{sec:orb_method}
\begin{figure*}
\centering
\includegraphics[scale=0.45]{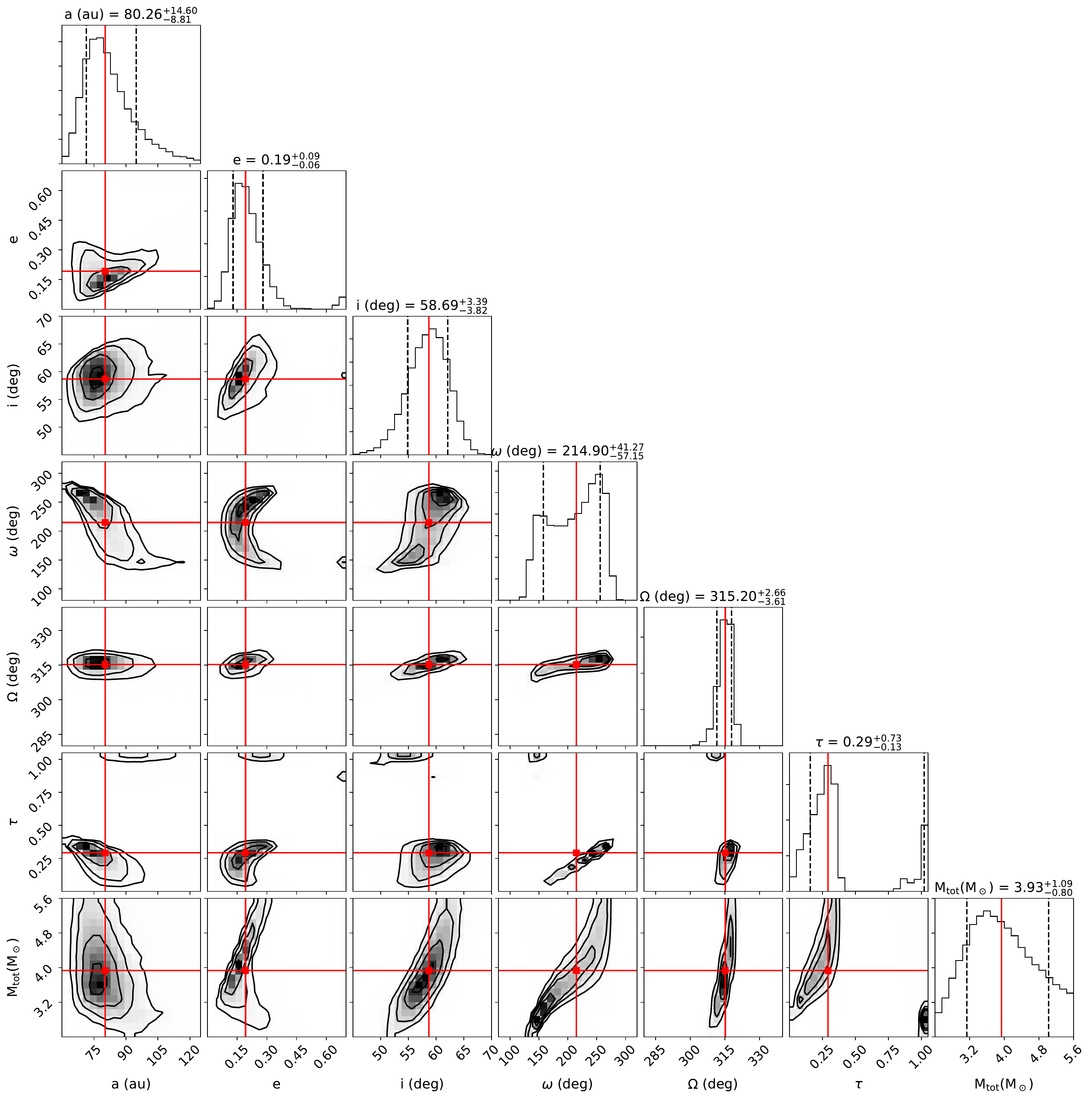}
\caption{Corner plot showing the posterior distribution of the orbital parameters from top to bottom: semi-major axis  in au, eccentricity $e$, inclination $i$ in degrees, argument of periastron $\omega$ in degrees, longitude of ascending node $\Omega$ in degrees, epoch of periastron passage $\tau$ (measured as fraction of orbit compared to a reference date, see~\ref{sec:orb_method}) and total mass $M_{tot}$ in \msun. The red lines indicate the median for each parameter and the dashed-lines correspond to symmetric confidence intervals around the median and enclosing 68\% of the probability. 
\label{fig:corner}}
\end{figure*}

We use the open-source software package orbitize! \citep{2020BluntOrbitize} which uses a parallel-tempered Markov Chain Monte Carlo (MCMC) algorithm (\citealt{2013ForemanMackeyemcee,2016VousdenDynamic}) to fit orbits using both positional and line of sight velocity observations (Table~\ref{table:los}). We input the twelve VLA positions and the single ALMA positions and radial velocity measurements. To assess possible priors, we first explore the parameter space with a grid fitting method \citep{2008KohlerOrbits}. The final implemented priors are  Gaussian with mean 100 au and standard deviation of 50 au for the semimajor axis $a$, uniform between 0 and 0.85 for eccentricity $e$, LogUniform with a minimum of 2 \msun\ and a maximum of 6 \msun for the total mass $M_{tot}$, uniform between $40\degr$ and $80\degr$ for the inclination $i$, uniform between 0$\degr$ and 360$\degr$ for both the argument of periastron $\omega$ and the longitude of ascending node $\Omega$, and Gaussian with a mean of 0.5 and standard deviation of 0.3 for the epoch of periastron passage $\tau$. The latter is expressed in orbitize! as a fraction of the orbital period past a specified reference date $t_{ref}$ (default January 1, 2020) and thus with possible values between 0 and 1. These priors are only weakly informed. We note that changing all priors to uniform within comparable ranges does not affect significantly our results. The distance was fixed to 141 pc. We ran 20,000 steps with 1,000 walkers per temperature with 20 temperatures. We removed 10,000 steps as burn-in. The resultant median values and confidence intervals from the posterior distribution are $a=80.26^{+14.60}_{-8.81}$ au, $e=0.19^{+0.09}_{-0.06}$, $i=58.69\degr^{+3.39}_{-3.82}$, $\omega=214.90\degr^{+41.27}_{-57.15}$, $\Omega=315.20\degr^{+2.66}_{-3.61}$, $\tau=0.29^{+0.73}_{-0.13}$, and $M_{tot}=3.93^{+1.09}_{-0.80}$ \msun. The full posterior distributions, medians and confidence intervals are displayed in Figure~\ref{fig:corner}.

\begin{deluxetable}{lccl}[ht]
\tablecolumns{4}
\tablecaption{Selected VLA observations for analysis of protostellar motion}             
\label{tb:vla_data}      
\centering                          
\tablehead{\colhead{Epoch} & \colhead{Frequency} & \colhead{Synthesized beam} & \colhead{References} \\
\colhead{year} & \colhead{(GHz)} & \colhead{} & \colhead{}}
\startdata
1987.7 &15 & 0.19\arcsec$\times$0.09\arcsec & (1,2,3) \\
1989.1 &8 & 0.34\arcsec$\times$0.19\arcsec& (4,2,3)\\
1989.1&	22& 	0.18\arcsec$\times$0.09\arcsec & (4,2,3)\\
1994.3&	8&	0.34\arcsec$\times$0.16\arcsec & (5,2,3)\\
2003.5&	43& 	0.09\arcsec$\times$0.05\arcsec & (2,3,6)\\
2003.7&	8& 	0.39\arcsec$\times$0.19\arcsec & (2,3,7)\\
2005.2&	43&	0.30\arcsec$\times$0.17\arcsec & (7,3)\\
2011.2&	41&	0.30\arcsec$\times$0.14\arcsec & (3) \\
2011.4&	41&	0.13\arcsec$\times$0.10\arcsec & (3) \\
2011.4&	41&	0.08\arcsec$\times$0.05\arcsec & (3) \\
2012.9&	33&	0.10\arcsec$\times$0.04\arcsec & (3) \\
2013.0&	41&	0.09\arcsec$\times$0.04\arcsec & (3) \\
\enddata
\tablerefs{
(1) \cite{1989WoottenDuplicity}; 
(2) \cite{2005ChandlerIRAS16293};
(3) \cite{2019HernandezGomezNature};
(4) \cite{1992MundyIRAS16293};
(5) \cite{2002LoinardLarge};
(6) \cite{2005RodriguezIRAS16293};
(7) \cite{2007LoinardNew}
}
\end{deluxetable}

\section{Bound analysis A-B}
Following the method in Section~\ref{sec:bound_analysis} we first use the most-recent published proper motions for B \citep{2019HernandezGomezNature} and previous determinations of its line-of-sight velocity \cite{2012PinedaFirst}, to estimate a minimum total mass A+B for the triple system (A1, A2 and B) to be bound. The proper motions alone show that A1 has moved significantly with respect to A2 and B ($5\pm1$ \kms in both cases). On the other hand, A2 has not moved significantly with respect to B ($1.5\pm1$ \kms), suggesting that the center of mass of A1-A2 is located closer to A2 (i.e., mass ratio A1/A2$<$1). Using a mass ratio between 0 and 1 and the analysis in Section~\ref{sec:bound_analysis}, results in a total minimum mass between $\sim$ 2-7 \msun, for IRAS 16293 to be a bound triple. Source B has a mass close to $~$1 \msun between its circumstellar disk and protostellar masses\footnote{We note that Source B is face-on and extremely optically thick within the recoverable scales of our data, thus we could not perform the same kinematic analysis as in Source A.} \citep{2012PinedaFirst,2018OyaSourceB}, and Source A has a combined mass of at least 1 \msun, while the mass in the large scale envelope around the three sources is about 4-6 \msun \citep{2018JacobsenALMAPILS,2020LadjelateHerschel}. Thus it is reasonable to conclude that B is also bound to A1 and A2, forming a hierarchical triple system.

\renewcommand{\thefigure}{F\arabic{figure}}
\renewcommand{\theHfigure}{F\arabic{figure}}
\setcounter{figure}{0}
\section{Additional figures}

\begin{figure*}[ht!]
\includegraphics[width=\textwidth]{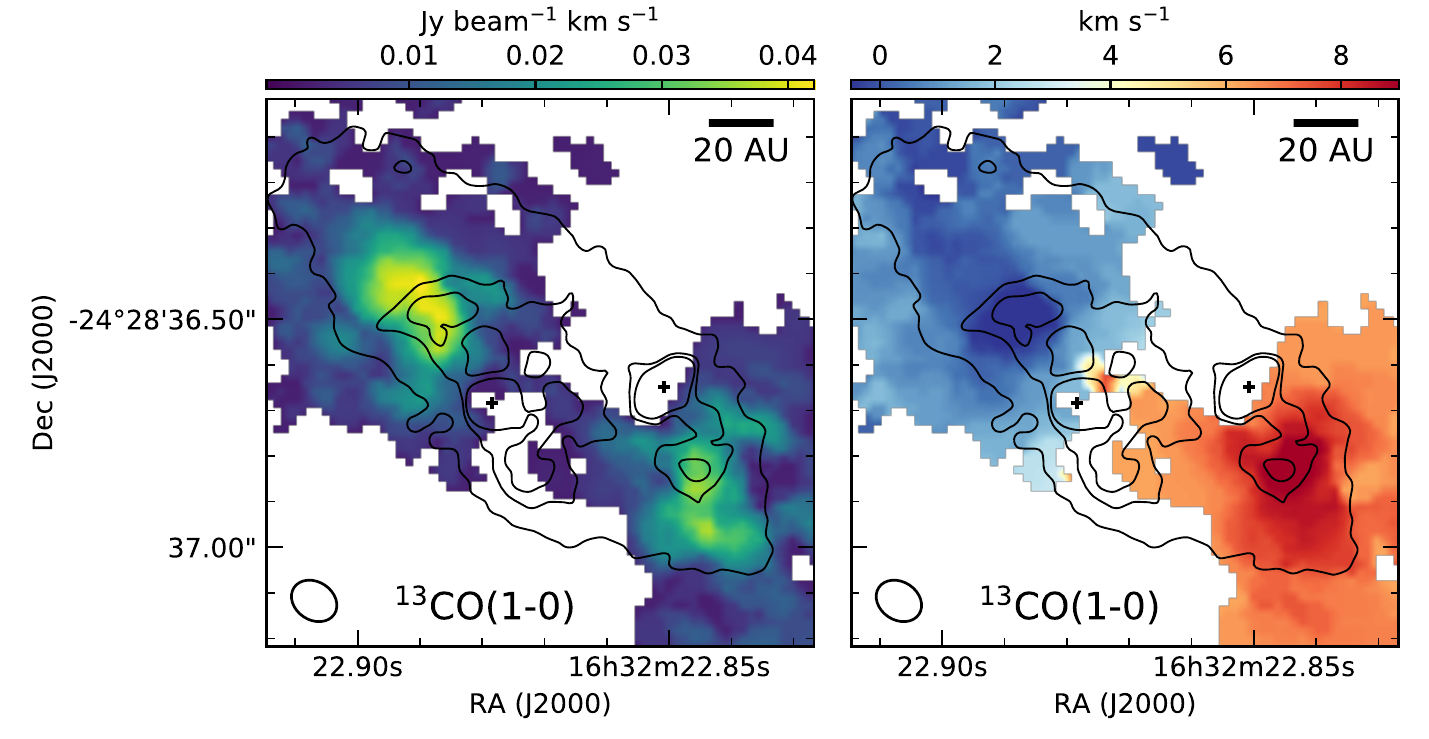}
\caption{$^{13}$CO (1-0) moment 0 (left) and 1 (right) maps towards IRAS 16293 A (color).  The velocity range for the $^{13}$CO (1-0) was split into two to avoid channels with artifacts due to missing flux. The two ranges were $[-7.02,2.86]$ \kms\ and $[6.28,12.74]$ \kms, same as CS (2-1) in Figures~\ref{fig:mom0} and~\ref{fig:mom1}. Black contours show the 3 mm continuum emission at levels 124 $\mu$Jy, 320 $\mu$Jy and 448 $\mu$Jy to identify the circumbinary disk-like structure, the compact sources A1 and A2 and the smaller scales substructures around them. Crosses mark the peak location for A1 and A2 (Table~\ref{tb:3mmfit}).  The beam is shown in the bottom left corner of each panel. 
\label{fig:13co_mom}}
\end{figure*}

\begin{figure*}[ht!]
\includegraphics[width=\textwidth]{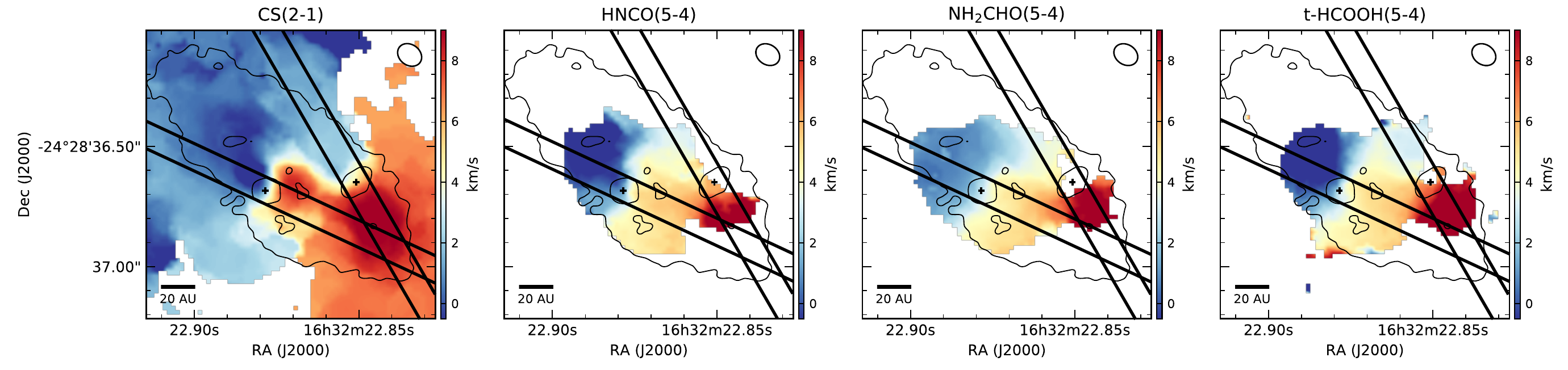}
\caption{Moment 1 maps towards IRAS 16293 A (color), same as Figure~\ref{fig:mom1}. The black straight lines show the direction and width of the position-velocity cuts used for the gas kinematic analysis (Sections~\ref{sec:los} and~\ref{sec:kep}). The P.A. of the cuts correspond to $65^{\circ}$ and $30^{\circ}$ for A1 and A2, respectively.\label{fig:cuts}}
\end{figure*}

\begin{figure*}[ht!]
\centering
\includegraphics[width=\textwidth]{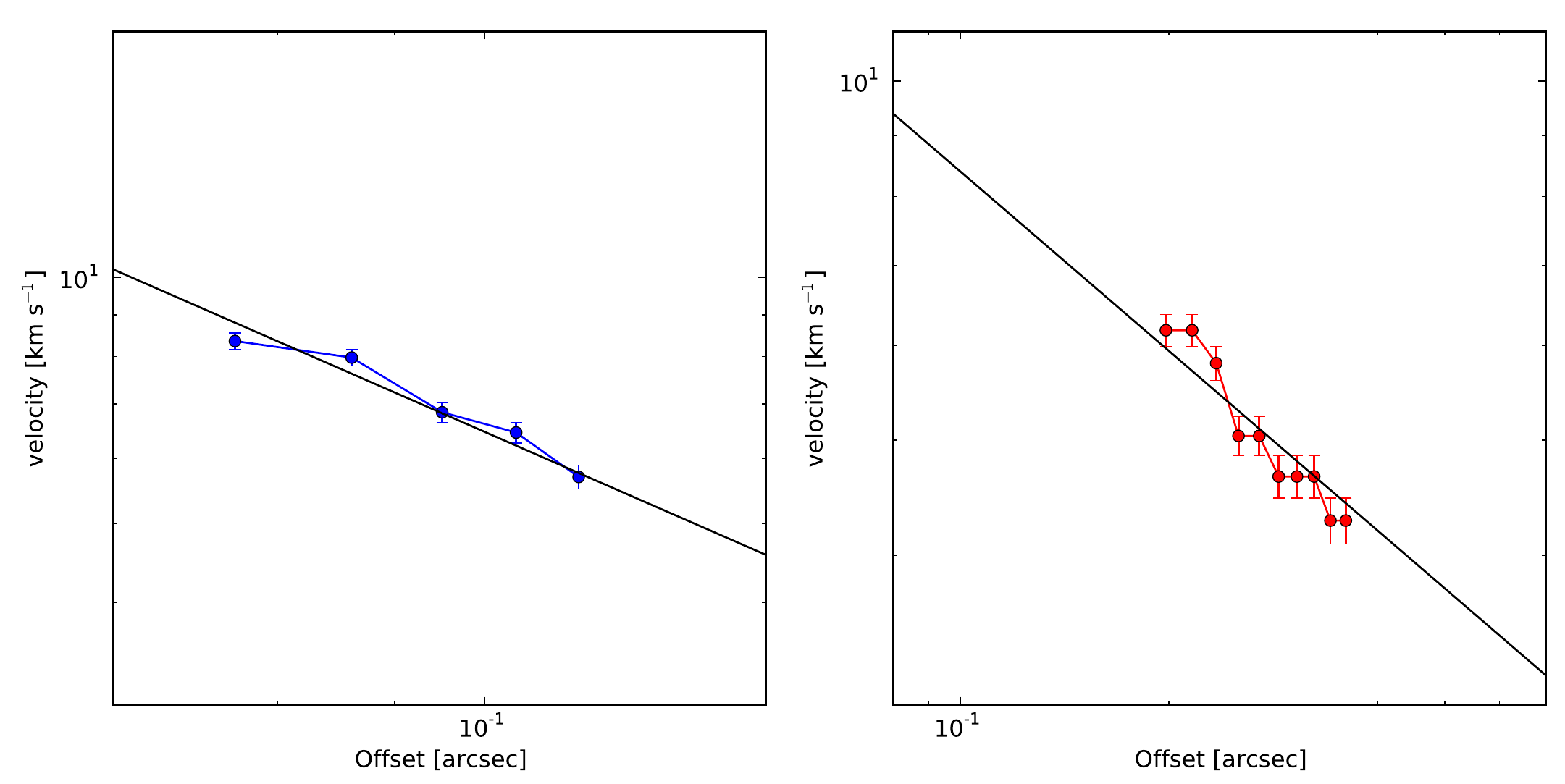}
\caption{CS (2-1) Velocity profiles extracted using the upper edge method \citep{2016SeifriedRevealing} for A1 (left) and A2 (right). The symbols corresponds to the data points in agreement with a Keplerian power law profile. Errors are equal to the channel width (0.38 \kms). The lines show the Keplerian power law fit to the data, resulting in the mass constrains discussed in Section~\ref{sec:kep}.
\label{fig:cs_upperedgefit}}
\end{figure*}


\bibliography{references}{}
\bibliographystyle{aasjournal}



\end{document}